\documentclass[aps,prl,twocolumn,superscriptaddress,nofootinbib]{revtex4-2}

\usepackage{latexsym}
\usepackage{amsmath}
\usepackage{amssymb}
\usepackage{amsfonts}
\usepackage{bm}
\usepackage{appendix}
\RequirePackage{lineno}

\usepackage{color}
\definecolor{purple}{rgb}{0.5,0,0.5}
\definecolor{blue}{rgb}{0.0,0,0.9}
\definecolor{prdblue}{rgb}{0.133,0.118,0.498}
\usepackage[colorlinks=true, pdfstartview=FitV, linkcolor=prdblue, citecolor= prdblue, urlcolor=prdblue]{hyperref}

\usepackage{supertabular} 
\usepackage{placeins}
\usepackage{epsfig}
\usepackage{graphicx}
\usepackage{hyperref}
\usepackage{comment}
\usepackage{physics}
\usepackage{multirow}
\usepackage{float}
\usepackage{dcolumn}
\usepackage{siunitx}

\hypersetup{
  breaklinks=true,
  colorlinks = true,
  linkcolor = blue,
  anchorcolor = blue,
  citecolor = blue,
  filecolor = blue,
  pagecolor = blue,
  urlcolor = blue
}
\hypersetup{
  bookmarks=true,
  bookmarksnumbered=true,
  bookmarkstype=toc,
  linktocpage=true
}

\newcommand{\lcp}{\Lambda_c^+}
\newcommand{\lcm}{\bar{\Lambda}_c^-}
\newcommand{\pip}{\pi^+}
\newcommand{\pim}{\pi^-}
\newcommand{\piz}{\pi^0}
\newcommand{\lmdetapi}{\Lambda \pi^+ \eta}
\newcommand{\sigmaeta}{\Sigma(1385)^+ \eta}
\newcommand{\lambdapi}{\Lambda(1670) \pi^+}
\newcommand{\lambdaa}{\Lambda a_0(980)^+}
\newcommand{\br}[1]{\mathcal{B}(#1)}

\def \dE {\Delta E}
\def \ebeam {E_{\rm{beam}}}
\def \mbc {M_{\rm{BC}}}

\def \gev  {\mbox{GeV}}

\def \gevcc{\mbox{GeV/$c^2$}}

\def \mevcc{\mbox{MeV/$c^2$}}

\def\sgm1385{\Sigma(1385)^+}
\def\lmd1670{\Lambda(1670)}
\def\a980{a_0(980)^+}

\def \romanOne   {\uppercase\expandafter{\romannumeral1}}
\def \romanTwo   {\uppercase\expandafter{\romannumeral2}}
\def \romanThree {\uppercase\expandafter{\romannumeral3}}
\def \romanFour  {\uppercase\expandafter{\romannumeral4}}
\def \romanFive  {\uppercase\expandafter{\romannumeral5}}
\def \romanSix   {\uppercase\expandafter{\romannumeral6}}
\def \romanSeven {\uppercase\expandafter{\romannumeral7}}
\def \romanEight {\uppercase\expandafter{\romannumeral8}}

\begin{document}


\setlength{\oddsidemargin}{-0.5cm} \addtolength{\topmargin}{15mm}

\title{\boldmath Observation of $\Lambda_c^+ \to \Lambda a_0(980)^+$ and Evidence for $\Sigma(1380)^+$ in $\Lambda_c^+ \to \Lambda \pi^+ \eta$}

\author{
M.~Ablikim$^{1}$, M.~N.~Achasov$^{4,c}$, P.~Adlarson$^{75}$, O.~Afedulidis$^{3}$, X.~C.~Ai$^{80}$, R.~Aliberti$^{35}$, A.~Amoroso$^{74A,74C}$, Q.~An$^{71,58,a}$, Y.~Bai$^{57}$, O.~Bakina$^{36}$
% ...
\\
\vspace{0.2cm}
(BESIII Collaboration)\\
\vspace{0.2cm} {\it
$^{1}$ Institute of High Energy Physics, Beijing 100049, People's Republic of China\\
% ...
$^{3}$ Bochum  Ruhr-University, D-44780 Bochum, Germany\\
$^{4}$ Budker Institute of Nuclear Physics SB RAS (BINP), Novosibirsk 630090, Russia\\
% ...
$^{35}$ Johannes Gutenberg University of Mainz, Johann-Joachim-Becher-Weg 45, D-55099 Mainz, Germany\\
$^{36}$ Joint Institute for Nuclear Research, 141980 Dubna, Moscow region, Russia\\
% ...
$^{57}$ Southeast University, Nanjing 211100, People's Republic of China\\
$^{58}$ State Key Laboratory of Particle Detection and Electronics, Beijing 100049, Hefei 230026, People's Republic of China\\
% ...
$^{71}$ University of Science and Technology of China, Hefei 230026, People's Republic of China\\
% ...
$^{74}$ University of Turin and INFN, (A)University of Turin, I-10125, Turin, Italy; (B)University of Eastern Piedmont, I-15121, Alessandria, Italy; (C)INFN, I-10125, Turin, Italy\\
$^{75}$ Uppsala University, Box 516, SE-75120 Uppsala, Sweden\\
% ...
$^{80}$ Zhengzhou University, Zhengzhou 450001, People's Republic of China\\
\vspace{0.2cm}
$^{a}$ Deceased\\
% ...
$^{c}$ Also at the Novosibirsk State University, Novosibirsk, 630090, Russia\\
% ...
}

}

\begin{abstract}
Based on $6.1~\mathrm{fb}^{-1}$ of $e^+e^-$ annihilation data collected at center-of-mass energies from 4.600 to 4.843~GeV with the BESIII detector at the BEPCII collider, a partial wave analysis of $\Lambda_c^+\to\Lambda\pi^+\eta$ is performed, and branching fractions and decay asymmetry parameters of intermediate processes are determined. The process $\Lambda_c^+\to\Lambda a_0(980)^+$ is observed for the first time, and evidence for the pentaquark candidate $\Sigma(1380)^+$ decaying into $\Lambda\pi^+$ is found with statistical significance larger than $3\sigma$ with mass and width fixed to theoretical predictions. The branching fraction product $\mathcal{B}[\Lambda_{c}^{+} \to \Lambda a_0(980)^+] \; \mathcal{B}[ a_0(980)^+ \to \pi^{+}\eta]$ is determined to be $(1.05 \pm 0.16_{\mathrm{stat}} \pm 0.05_{\mathrm{syst}}  \pm 0.07_{\mathrm{ext}})\%$, which is larger than theoretical calculations by 1-2 orders of magnitude. Here the third (external) systematic is from $\mathcal{B}(\Lambda_{c}^{+} \to \Lambda \pi^+ \eta)$. Finally, we precisely obtain the absolute branching fraction $\mathcal{B}(\Lambda_{c}^{+} \to \Lambda \pi^+ \eta) = (1.94 \pm 0.07_{\mathrm{stat}} \pm 0.11_{\mathrm{syst}})$\%.
\end{abstract}

\maketitle

Apart from the traditional bound states like mesons and baryons, the quark model~\cite{QM_GellMann,QM_Zweig} allows for  more complex structures such as tetraquarks, pentaquarks, hybrids, glueballs, and hadronic molecular states. For the studies of these exotic states, many important achievements have been made~\cite{XYZ_review1,XYZ_review2,XYZ_review3,XYZ_review4,XYZ_review5,XYZ_review6,XYZ_review7}, especially in quarkonium, $D$ meson, and $B$ meson decays. Studies of baryon decays are relatively rare, except for studies of $\Lambda_b^0$ decays into $P_c$ or $X(3872)$ states by the LHCb experiment~\cite{LHCb:2019kea,LHCb:2015yax,LHCb:2019imv}. Replacing the $b$ quark with a $c$ quark, the much lighter charm baryons lie at the boundary between perturbative and nonperturbative regions. Given the very interesting results already achieved in the first studies of heavy exotic states in bottom baryon decays, studies using charm baryon decays provide a new and exciting opportunity to probe lighter exotic states. 

The exact nature of the scalar meson $\a980$ remains elusive, with various interpretations proposed. 
These include a conventional $q\bar{q}$ meson~\cite{Soni:2020sgn,Klempt:2021nuf}, a compact tetraquark~\cite{Achasov:1999wv,Jaffe:2004ph}, a superposition of both~\cite{Alexandrou:2017itd}, or a dynamically generated threshold effect~\cite{Weinstein:1982gc,Janssen:1994wn,Oller:1997ti,Sekihara:2014qxa,Ahmed:2020kmp}. Reference~\cite{Sharma:2009zze} adopted the compact tetraquark assumption to study the $\lcp\to\lambdaa$ decay, as the $q\bar{q}$ picture failed to explain the measured $\mathcal{B}[\lcp\to p f_0(980)]$~\cite{ACCMOR:1990gke}, where the $f_0(980)$ is regarded as the scalar octet partner of $\a980$ in the $q\bar{q}$ model. The $\lcp\to\lambdaa$ branching fraction (BF) was calculated to be $1.9\times10^{-4}$ based on factorization and the pole model, where the pole term was found to dominate over factorizable contributions.
In a different perspective, Ref.~\cite{Yu:2020vlt} proposed a significant enhancement of the BF to $(1.7^{+2.8}_{-1.0}\pm0.3)\times 10^{-3}$ by considering the process $\lcp \to \sigmaeta$, followed by rescattering $\sigmaeta \to \lambdaa$. Here, the calculated $\mathcal{B}[\lcp \to \sigmaeta]$ in the topological scheme~\cite{Hsiao:2020iwc} is employed as an input. Contributions from other processes, such as $\lcp \to \lambdapi \to \lambdaa$ and the triangle singularity enhanced $\lcp \to \Sigma^{*}\eta(N^{*}\bar{K}^0) \to \lambdaa$, are estimated to be less than $1\times10^{-3}$.
Moreover, due to the proximity of the $\a980$ pole mass to the $K\bar{K}$ threshold, the $\a980$ line shape exhibits a distinct cusp structure, a characteristic feature indicative of its molecular nature~\cite{Wang:2022nac}.
Therefore, the $\lcp\to\lmdetapi$ decay provides a good platform to study the internal structure of $\a980$.

The study of low-lying excited baryons with $J^P=1/2^{-}$ is crucial in hadron physics~\cite{Wang:2024jyk}. Historically, to address the reverse mass-order reverse of the $N(1535)$ and $\Lambda(1405)$ states, theorists proposed the pentaquark model~\cite{Helminen:2000jb,Zhang:2004xt,Zou:2007mk} and the meson cloud and molecular model~\cite{Kaiser:1995cy,Jido:2003cb}. These models predict the lowest $\Sigma^*_{1/2^-}$ resonance around 1380 MeV/$c$\(^2\)~\cite{Wu:2009tu}, close to the $N\bar{K}$ mass threshold~\cite{Khemchandani:2011mf}. Experimental and theoretical investigations on $\Sigma(1380)^+$ as well as other light pentaquark states containing strange quarks have been conducted in various processes~\cite{Wu:2009tu,Wu:2009nw,Gao:2010ve,Khemchandani:2018amu,Pan:1969bq,Pan:1969ad,ZEUS:2004lje,Zychor:2005sj,Gao:2010hy,CLAS:2013rjt,Chen:2013vxa,Roca:2013cca,Kim:2021wov,Lyu:2023oqn,Zou:2007mk,Wang:2015qta,Liu:2017hdx,Belle:2022ywa,Dey:2014tfa,Lebed:2015dca,Belle:2017tfw,Xie:2017mbe,Liu:2018nse,Xie:2020jlz}. However, establishing the lowest $\Sigma^*_{1/2^-}$ resonance remains a challenge. The $\lcp\to\lmdetapi$ decay has been highlighted as a golden channel~\cite{Xie:2017xwx,Lyu:2024qgc}. The $\Lambda\pip$ mode, representing a pure $I=1$ combination, excludes influences from $\Lambda^*$ resonances as compared to the $\Sigma\pi$ and $pK$ modes. Also, the influences from the $\Sigma(1385)^+$ and $\Lambda(1670)$~\cite{BESIII:2018qyg,Belle:2020xku} on the $\Sigma(1380)^+$ can be distinguished. This is because $\Lambda(1670)$ predominantly affects the high end of the $M(\Lambda\pip)$ spectrum, while the $\Sigma(1385)^+$ exhibits a different spin-parity resulting in a distinct angular distribution.

In this Letter, the first partial wave analysis (PWA) of the $\lcp \to \lmdetapi$ decay is performed by using 11 datasets at center-of-mass (c.m.) energies from 4.600 to 4.843~GeV~\cite{BESIII:2015qfd,BESIII:2015zbz,BESIII:2022dxl,BESIII:2022ulv,Ke:2023qzc}, where $\lcp$ is dominantly produced via pair production $e^+e^-\to\lcp\lcm$. There is no sufficient energy for producing additional hadrons below 4.7~GeV, and the process $e^+e^-\to\lcp\lcm\pi^0$ is highly suppressed between 4.7 and 4.843~GeV. The datasets used are accumulated with the BESIII detector at the BEPCII collider and correspond to an integrated luminosity of $6.1~\mathrm{fb}^{-1}$. Detailed information about BESIII and BEPCII can be found in Refs.~\cite{Yu:IPAC2016-TUYA01, Ablikim:2009aa, Ablikim:2019hff,detvis}. The simulated ``inclusive Monte Carlo (MC) sample'' is described in Ref.~\cite{BESIII:2022udq}. In the ``phase-space (PHSP) signal MC sample'' and the ``PWA signal MC sample'', $\lcp\to\lmdetapi$ decays are simulated with a uniform PHSP distribution and our PWA result, respectively, while the $\lcm$ decays inclusively. Throughout this Letter, charge-conjugate modes are implied unless explicitly noted.

We use a single-tag (ST) method~\cite{Li:2021iwf}, where the $\lcp$ is reconstructed via the cascade decays $\Lambda_c^+ \to \Lambda \pi^+ \eta, \Lambda \to p \pi^-, \eta \to \gamma \gamma$ and $\eta \to \pip \pim \piz$, $\piz \to \gamma \gamma$. The requirements for selecting charged tracks, photon showers, and particle identification (PID) for the proton and pion follow the previous BESIII analysis~\cite{BESIII:2022udq}. To reconstruct $\Lambda$ candidates, the $p\pim$ pairs are constrained to originate from a common vertex by requiring the $\chi^2$ of a vertex fit to be less than 100 and the $p\pim$ invariant mass to satisfy $1.08<M_{p\pim}<1.15\,\gevcc$. To reconstruct $\eta, \piz \to\gamma\gamma$ candidates, the $\gamma\gamma$ invariant mass $M_{\gamma\gamma}$ is required to be within $[0.500,0.600]\,\gevcc$ ($[0.105,0.150]\,\gevcc$). To improve the momentum resolution, a one-constraint kinematic fit is performed by constraining $M_{\gamma\gamma}$ to the known $\eta, \piz$ masses~\cite{pdg2023}, and the fit $\chi^2$ must be less than 20 (200). The updated momenta are used in further analysis. To reconstruct $\eta \to \pip\pim\piz$ candidates, the $\pip\pim\piz$ invariant mass, $M_{\pip\pim\piz}$, is required to be within $(0.500,0.600)\,\gevcc$. If there are multiple $\Lambda_{c}^+$ combinations in an event, we choose the candidate with the minimum magnitude of the energy difference, defined as $\dE \equiv E_{\Lambda_c}-\ebeam $, where $E_{\Lambda_c}$ is the energy of the detected $\lcp$ candidate in the $e^+e^-$ rest frame, and $\ebeam$ is the beam energy. Furthermore, the requirement $-0.1<\dE<0.1\,\gev$ is imposed.

To further suppress the backgrounds, a boosted decision tree with gradient boosting (BDTG)~\cite{hastie2009} based on the {\sc TMVA} package~\cite{Voss:2007jxm} is used. The input variables are $\dE$, $M_{p\pim}$, the ratio of the $\Lambda$ decay length to its uncertainty $L/\sigma_L$, $M_{\gamma\gamma}$, $M_{\pip\pim\piz}$ (\text{only for $\eta\to\pip\pim\piz$ channel}), the cosine of the helicity angle of $\eta, \piz \to\gamma\gamma$ decay, $\cos\theta_{\eta(\piz)}$, and the lateral moments of the showers with higher and lower energies $\mathrm{Lat}(\gamma_{\mathrm{High}})$ and $\mathrm{Lat}(\gamma_{\mathrm{Low}})$. The inclusive MC sample is input as the training set, in which the signal and nonsignal processes are tagged as signal and background, respectively. The resultant BDTG scores are required to be greater than 0.95 and 0.97 for the $\eta\to\gamma\gamma$ and $\eta\to\pip\pim\piz$ channels, respectively, chosen by optimizing the figure-of-merit $\mathrm{FOM}=\frac{S}{\sqrt{S+B}} \, \frac{S}{S+B}$. Here, $S$ $(B)$ is the number of  signal (background) events in the inclusive MC sample whose luminosity is normalized to the data.

An extended unbinned maximum likelihood fit is performed on the beam-constrained mass, $M_{\mathrm{BC}}=\sqrt{E_{\mathrm{beam}}^2-|\vec{p}|^2}$, distribution~\cite{Li:2021iwf} of each energy point to obtain the signal yields and purity in data, where $\vec{p}$ is the three-momentum of the ST $\Lambda_c^+$ candidate and $E_{\mathrm{beam}}$ is the beam energy, both evaluated in the $e^+ e^-$ center-of-mass system. The method is the same as Ref.~\cite{BESIII:2022udq}, and $1312 \pm 45$ signal events are obtained with purity of about 80\% in the signal regions, as shown in Supplemental Material~\cite{supple}. The result of the fit to the $M_{\rm BC}$ distribution from the combined $\eta\to\gamma\gamma$ and $\eta\to\pip\pim\piz$ channels at 4.682~GeV is shown in Fig.~\ref{fig:mbc_fit}, and the results at other energy points are shown in Supplemental Material~\cite{supple}. The event-by-event sWeight factor is calculated by the sPlot method~\cite{Pivk:2004ty}, according to the fit results. The sPlot method is a statistical tool dedicated to the exploration of data samples populated by several sources of events, e.g., signal and background. The sWeight factor as a function of discriminating variable like $M_{\mathrm{BC}}$ is designed such that it is normal to signal distribution but orthogonal to background distribution. After applying the sWeight factor, background does not contribute to the extracted signal distribution. In order to improve the momentum resolution, an additional three-constraint kinematic fit is applied, in which the $\lmdetapi$ invariant mass and the recoiled $\lcm$ mass are constrained to the known $\lcp$ mass, and the $p\pim$ invariant mass is constrained to the known $\Lambda$ mass~\cite{pdg2023}. The recoiled $\lcm$ momentum is calculated with the momentum of the initial $e^+e^-$ system and $\lmdetapi$ momentum. The updated momenta of the signal candidates from the kinematic fit are used in the PWA.

\begin{figure}[htbp]
	\centering
    \includegraphics[width=0.4\textwidth]{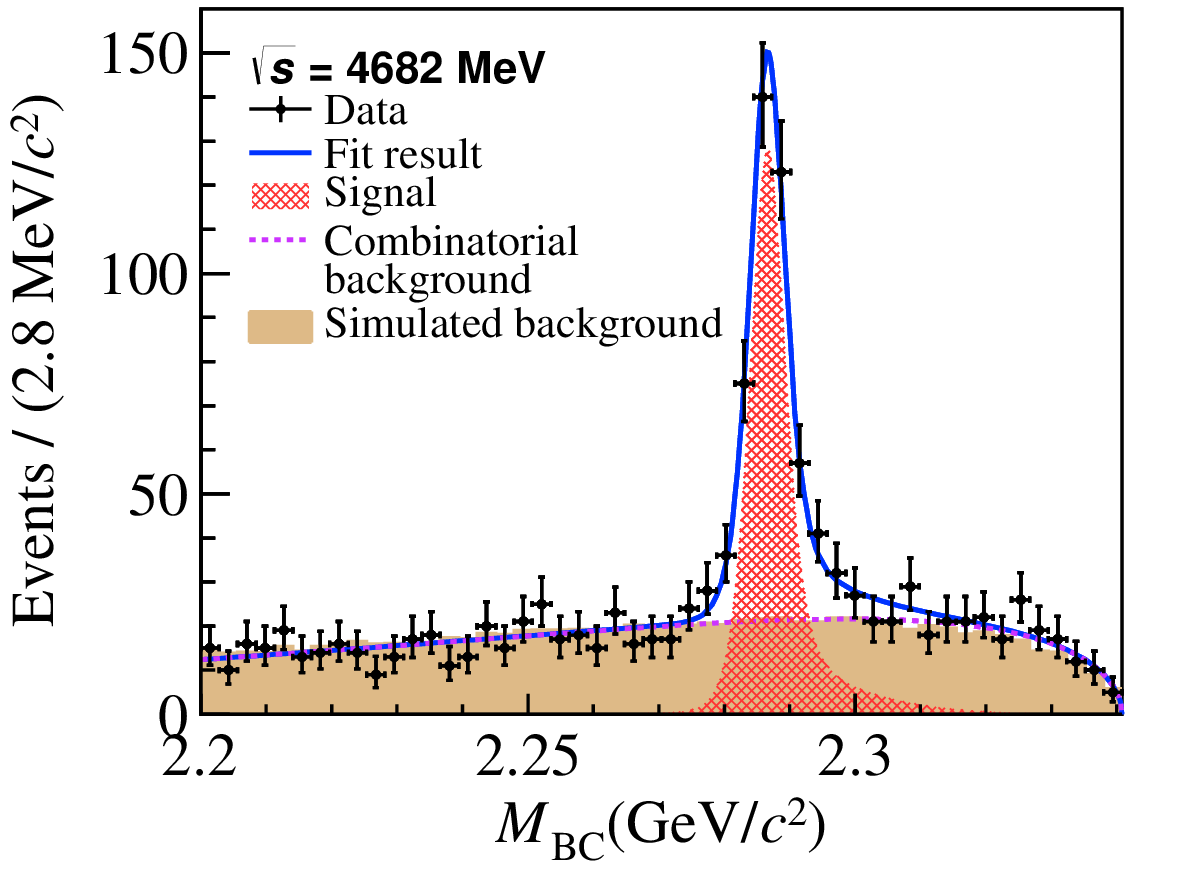}
	\caption{The fit to the $M_{\rm BC}$ distribution combined from the $\eta\to\gamma\gamma$ and $\eta\to\pi^+\pi^-\pi^0$ channels at 4.682~GeV. The points with error bars are data, the brown solid histogram is MC-simulated background, the red hatched histogram is signal, the violet dashed line is background shape, and the blue line is total fit.\label{fig:mbc_fit}}
\end{figure}

In the framework of the helicity amplitude formalism~\cite{Chung:1971ri,Richman:1984gh}, a PWA is performed by using the open-source framework {\sc TF-PWA}~\cite{tfpwa}. The fundamental concepts follow Ref.~\cite{BESIII:2022udq}. In this Letter, the amplitude is defined in the $e^+e^-$ rest frame. The parameters describing the amplitude of the $\lcp\to\lmdetapi$ decay are shared for each energy point. Moreover, the parameters describing the amplitude of the $\lcm$ decay are related to those of $\lcp$ via a parity transformation on the $\lcm$ candidates, under the assumption of $CP$ conservation. The $\lcp$ polarization components are fixed to $P_z=P_x=0$, $P_y(\theta_{ee},\alpha_0,\Delta_0)\propto \sqrt{1-\alpha_0^2}\sin\theta_{ee}\cos\theta_{ee}\sin\Delta_0$~\cite{Chen:2019hqi}. Here, $\theta_{ee}$ is the polar angle of the $\lcp$ with respect to the $e^+$ beam in the $e^+e^-$ c.m.~system, $\alpha_0$ is fixed to the values from Refs.~\cite{BESIII:2017kqg,BESIII:2023rwv}, and $\Delta_0$ is fixed according to polarization results in data. The decay amplitudes of the $\lcp$ decay are described with sequential helicity amplitudes for cascade quasi-two-body decay and the propagators of intermediate states. For decay chains with resonant intermediate states, the barrier factor term is included. For those with nonresonant (NR) intermediate states, the barrier factor term is omitted.

In the decay amplitude of $\lcp\to\lambdaa$, $\a980\to\pip\eta$, the propagator of the $\a980$ is described by the two coupled-channel Flatt\'{e} model~\cite{Flatte:1976xu}. The nominal mass and coupling constants of the $\a980$ decaying to the $\eta\pi$ and $K\bar{K}$ coupled channels are quoted from Ref.~\cite{BESIII:2016tqo}. For the NR decay, the dynamical function is set to be unity. In the decay chains of $\lcp\to\sigmaeta$, $\Sigma(1385)^{+}\to\Lambda\pip$ and $\lcp\to\lambdapi$, $\Lambda(1670)\to\Lambda\eta$, the relativistic Breit-Wigner (RBW) formula~\cite{BESIII:2022udq} is used as the propagator of the $\Sigma(1385)^{+}$ and $\Lambda(1670)$. The nominal mass and width of the $\sgm1385$ are fixed to the corresponding values from the Particle Data Group (PDG)~\cite{pdg2023}, and those of $\lmd1670$ are taken from a recent measurement~\cite{Sarantsev:2019xxm}. The amplitude of $\Lambda \to p\pim$ is constrained according to the decay asymmetry $\alpha_\Lambda$ from the PDG~\cite{pdg2023}. The full amplitude is the coherent sum of amplitudes of all decay chains, and the alignment $D$ functions are considered to align the helicities of the final state protons~\cite{BESIII:2018qyg,Wang:2020giv}. The construction of the signal probability density function and the derivations of fit fractions (FFs), interference, and corresponding statistical uncertainties follow the previous BESIII analysis~\cite{BESIII:2022udq}. The negative log-likelihood (NLL) is a sum over of all signal candidates considering the sWeight factor $w_i$ of $i$th event, $-\ln L=-a \sum_{i\in\mathrm{data}} w_i \ln P(p_i)$ with the normalization factor $a = \sum_{i\in\mathrm{data}} w_i \,/\, \sum_{i\in\mathrm{data}} w_i^2$~\cite{Langenbruch:2019nwe}.

\begin{figure*}[htbp]
	\centering
	\includegraphics[width=0.32\textwidth]{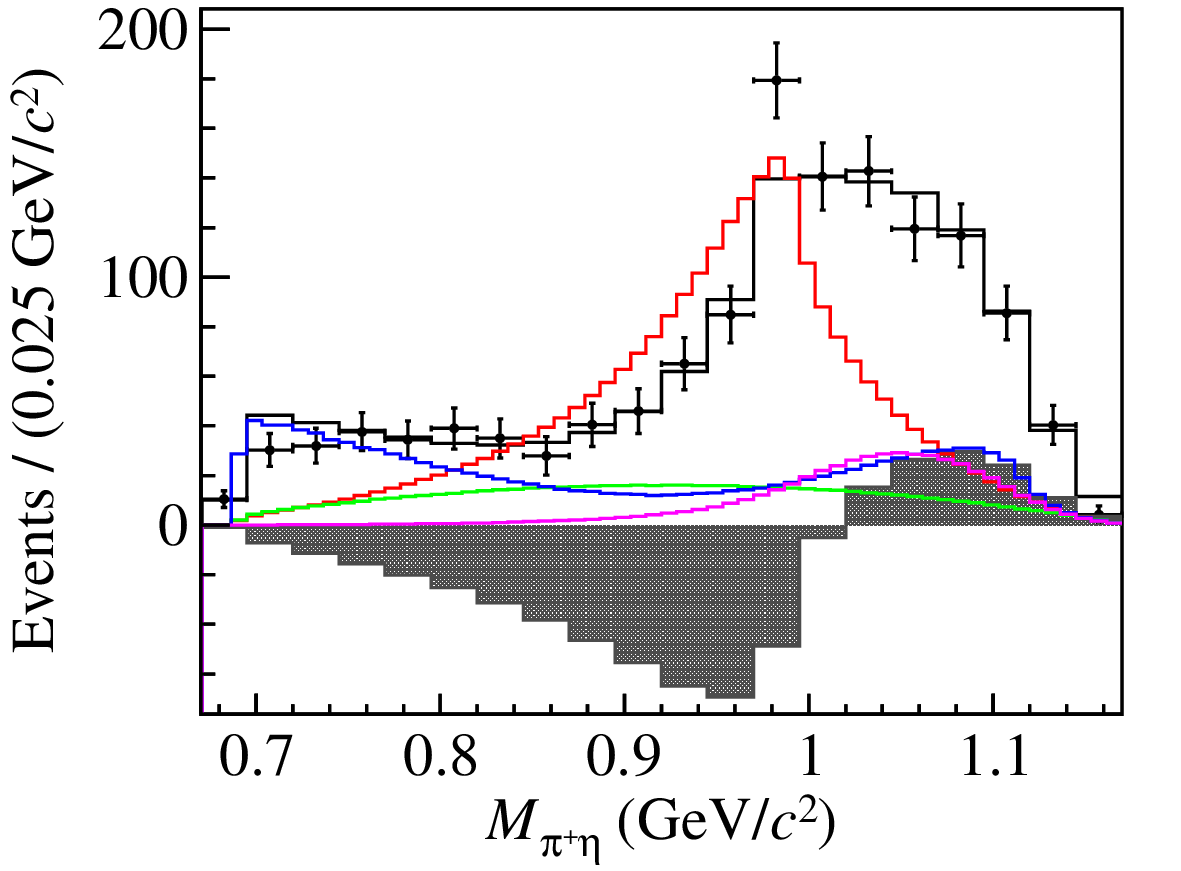}
	\includegraphics[width=0.32\textwidth]{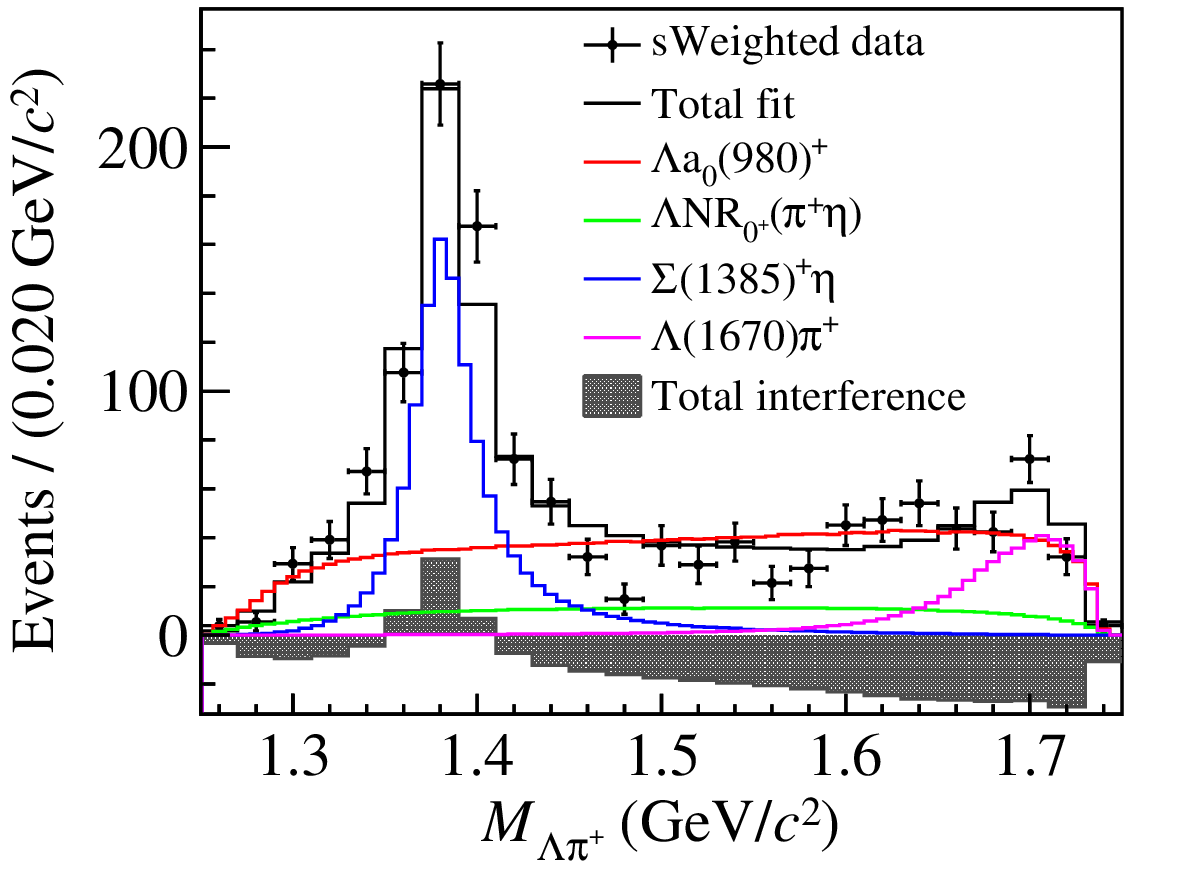}
	\includegraphics[width=0.32\textwidth]{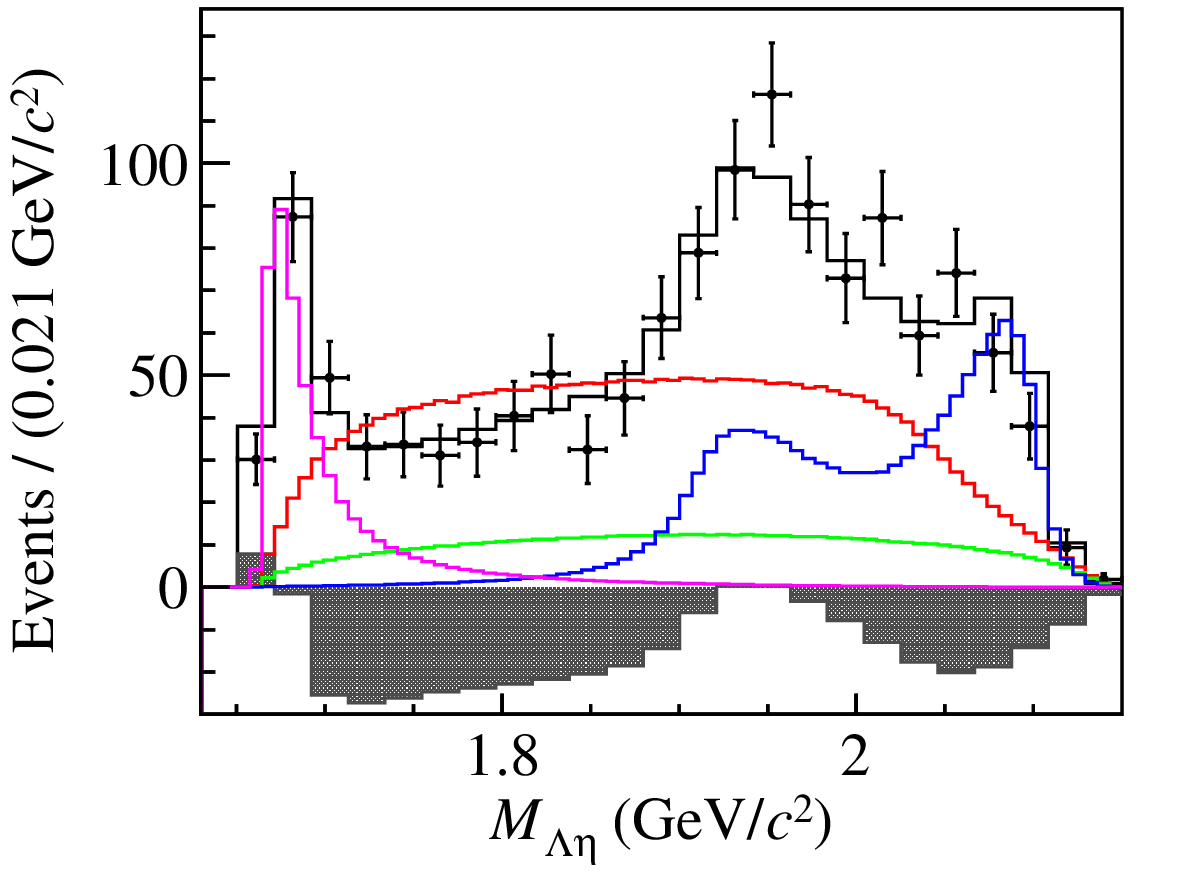}
	\caption{Projections of the fit results in the $M_{\pip\eta}$, $M_{\Lambda\pip}$, and $M_{\Lambda\eta}$ spectra. Points with error bars are sWeighted data at all energy points. The curves in different colors are different components.\label{fig:fit_mass}}
\end{figure*}

To determine the baseline solution of PWA, significant resonances $\sgm1385$, $\lmd1670$, and $\a980$ are added in the first trial. In the second iteration, other possible components are added one by one. The $\mathcal{S}$ wave $\pip\eta$ NR component NR$_{0^{+}}$ with highest significance is chosen. In the third iteration, the statistical significances of these amplitudes are all greater than $5\sigma$, as shown in Table~\ref{tab:finalresults}, and no other resonant component exceeds this threshold. The statistical significance is calculated from the change of the NLL values with and without including the component, taking into account the change of the number of degrees of freedom (d.o.f.). The fit results projected on different mass spectra are shown in Fig.~\ref{fig:fit_mass}. The fit results for the FFs and decay asymmetry parameters are listed in Table~\ref{tab:finalresults}, where the decay asymmetry parameter arises from the interference between partial wave amplitudes. Using fits to samples from pseudo-experiments, each matched to the data statistics, the pull distribution of each parameter is obtained. We correct the central value and scale the statistical uncertainty for a parameter if its pull distribution deviates significantly from the normal distribution. Since the fitted $\alpha_{\Lambda\a980}$ value is very close to its physical limit, an asymmetric statistical uncertainty is derived by scanning the NLL.

\begin{table}[htbp]
\begin{center}
\caption{Fit fractions, statistical significances $\mathcal S$, and decay asymmetry parameters $\alpha$ for different components in the baseline solution. The total FF is 113.9\%. The first uncertainty is statistical, and the second is systematic.} 
\small
	$\begin{array}{cccc}
	\hline \hline 
	\text { Process } & \text{FF (\%)} & \mathcal S & \alpha \\
	\hline 
	\Lambda\a980 & 54.0\pm8.4\pm2.6 &13.1\sigma & -0.91^{+0.18}_{-0.09}\pm0.08 \\
	\sgm1385\eta & 30.4\pm2.6\pm0.7 & 22.5\sigma & -0.61\pm0.15\pm0.04 \\
	\lmd1670\pip & 14.1\pm2.8\pm1.2 & 11.7\sigma & 0.21\pm0.27\pm0.33 \\
	\Lambda N\!R_{0^+} & 15.4\pm5.3 & 6.7\sigma & ... \\
	\hline \hline
	\end{array}$
\label{tab:finalresults}
\end{center}
\end{table}

\begin{figure*}[htbp]
	\centering
	\includegraphics[width=0.32\textwidth]{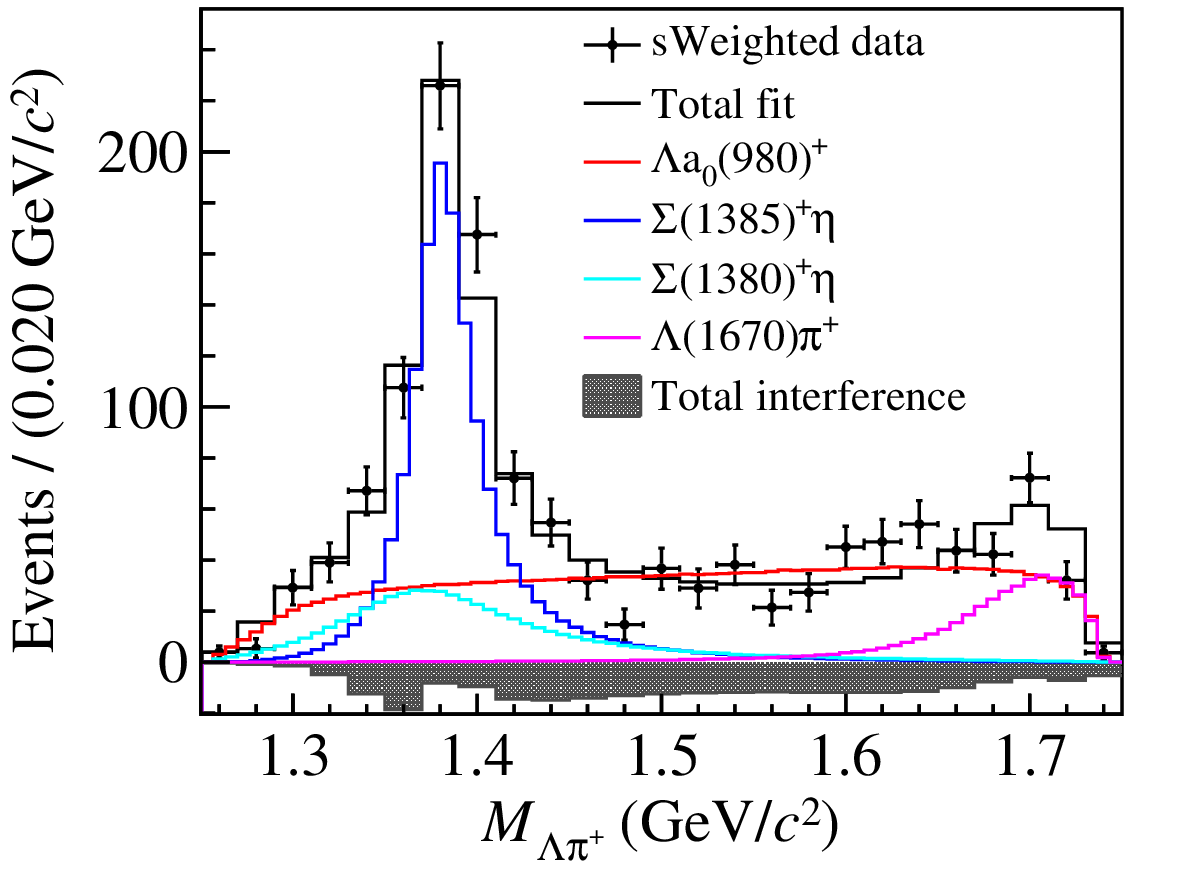}
    \includegraphics[width=0.32\textwidth]{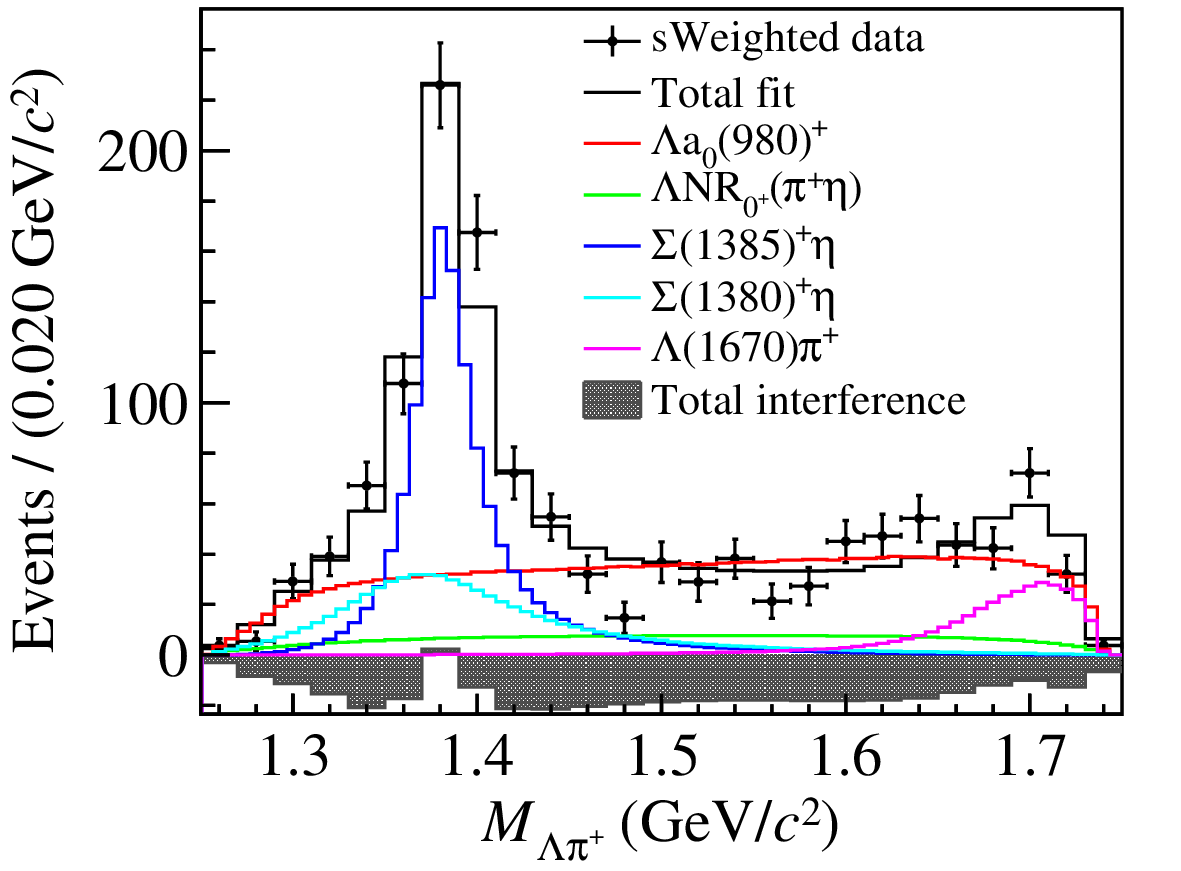}
    \includegraphics[width=0.32\textwidth]{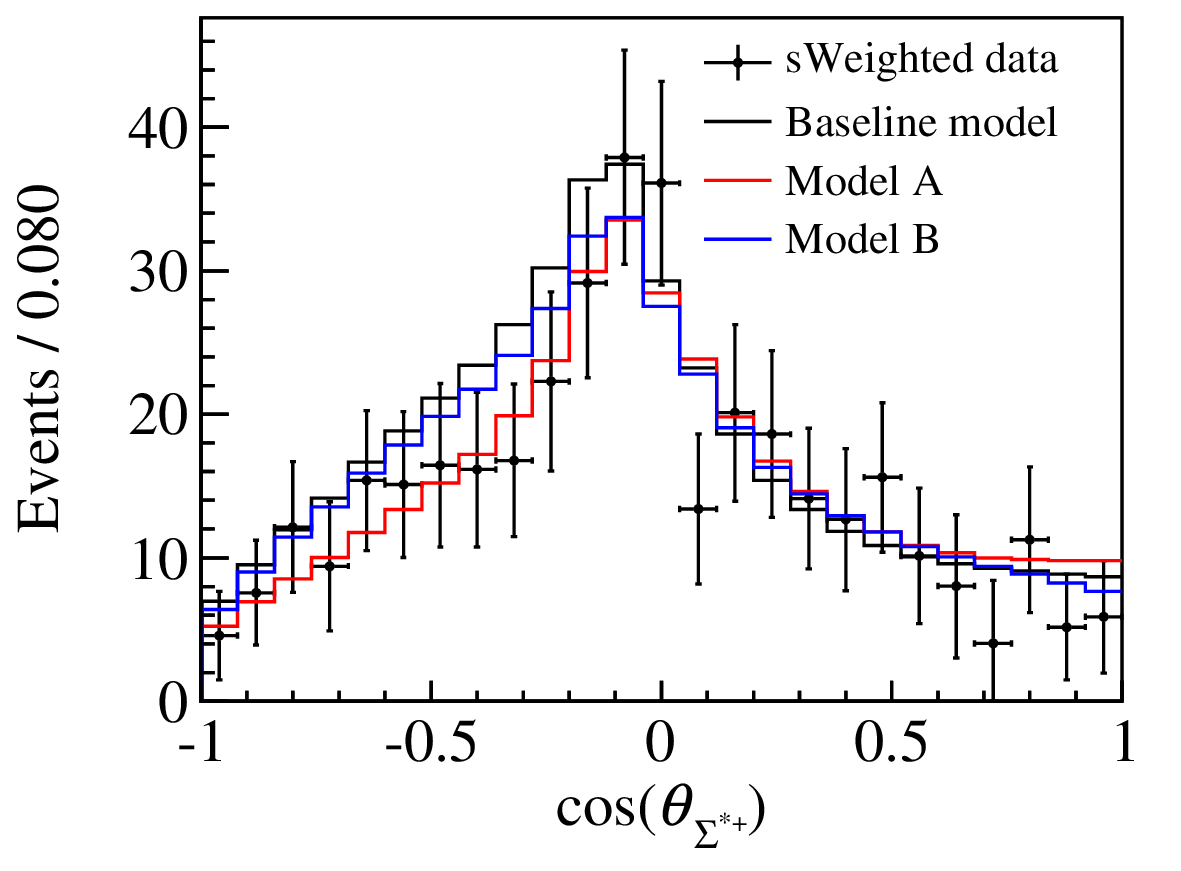}
	\caption{Left/middle: projections of models A and B on the $M_{\Lambda\pip}$ spectrum, respectively, where points with error bars are sWeighted data, and the curves in different colors are different components. Right: projections of the baseline model, models A and B on the $\Sigma^{*+}$ helicity angle, $\cos\theta_{\Sigma^{*+}}$, with the curve indicating the total fit.\label{fig:fit_mass_1380}}
\end{figure*}

Adopting the Breit-Wigner mass and width values of 1380~$\mevcc$ and 120~MeV, respectively, as predicted in Refs.~\cite{Xie:2017xwx,Wu:2009tu}, the potential pentaquark state $\Sigma(1380)^+$ is investigated in the signal process. In the construction of the baseline solution, NR$_{0^+}$ is introduced to better describe data with statistical significance of $6.7\sigma$, while that of the $\Sigma(1380)^+$ is slightly lower. To investigate the statistical significance of $\Sigma(1380)^+$, we construct ``model A'' [$\Lambda\a980$, $\sgm1385\eta$, $\lmd1670\pip$, $\Sigma(1380)^{+}\eta$] and `model B'' [$\Lambda\a980$, $\sgm1385\eta$, $\lmd1670\pip$, $\Sigma(1380)^{+}\eta$, $\Lambda \mathrm{NR}_{0^+}$]. Comparing model A (model B) with and without $\Sigma(1380)^{+}$, the statistical significance is determined to be $6.1\sigma$ ($3.3\sigma$) under model assumption of mass and width fixed to Refs.~\cite{Xie:2017xwx,Wu:2009tu}. 
The change in NLL is 24.1 and 9.2 for model A and model B, respectively, while number of d.o.f. changes are both 4. 
Projections onto the $M_{\Lambda\pip}$ spectrum for models A and B are illustrated in Fig.~\ref{fig:fit_mass_1380} left and middle, while the corresponding results for the FFs are detailed in Table~\ref{tab:FFwith1380}. Despite the overall significance of NR$_{0^+}$ being higher than that of $\Sigma(1380)^+$, a subtle preference for $\Sigma(1380)^+$ over NR$_{0^+}$ is discerned from the $\Sigma^{*+}$ helicity angle distribution in the $\a980$ signal region, $M_{\Lambda\pip}>1.44\,\gevcc$ and $M_{\Lambda\eta}>1.72\,\gevcc$. The comparison plot is shown in Fig.~\ref{fig:fit_mass_1380} right, and more details can be found in Supplemental Material~\cite{supple}. Additionally, various models are tested by replacing NR$_{0^+}$ with other excited states such as $\Sigma^{*+}$, $\Lambda^{*}$, $a_0^+$, and $a_2^+$, while considering systematic uncertainties arising from fixed mass and width parameters by varying them within $\pm1\sigma$~\cite{Wu:2009tu}, or float mass and width parameters. In all cases, the calculated statistical significances exceed $3\sigma$. Consequently, this study presents the first evidence for the $\Sigma(1380)^+$.

\begin{table}[htbp]
\begin{center}
\caption{Fit results of FFs and statistical significances for different components in alternative models including $\Sigma(1380)^{+}$. The total FFs are 115.8\% and 119.8\% for models A and B, respectively. The uncertainties are statistical only.}
\normalsize
	$\begin{array}{ccc}
	\hline \hline 
    \text { Process } & \text{Model A} & \text{Model B} \\
    \hline
    \Lambda\a980 & 52.9\pm4.5 \,(13.4\sigma) & 50.6\pm8.0 \,(11.1\sigma)\\
    \sgm1385\eta & 36.6\pm2.6 \,(15.8\sigma) & 31.3\pm3.0 \,(14.6\sigma)\\
    \lmd1670\pip & 10.7\pm1.4 \,(15.0\sigma) & 9.0\pm1.6 \,(11.9\sigma)\\
    \Sigma(1380)^{+}\eta & 15.5\pm4.4 \,(6.1\sigma) & 17.7\pm5.7 \,(3.3\sigma)\\
    \Lambda N\!R_{0^+} & ... & 11.3\pm4.4 \,(4.2\sigma)\\
	\hline \hline
	\end{array}$
\label{tab:FFwith1380}
\end{center}
\end{table}

The line shapes of the $\a980$ and $\lmd1670$ are also tested with the Final-State-Interaction (FSI) model~\cite{Wang:2022nac}, and alternative PWA fits are performed. No significant differences are observed in the results of the RBW and FSI models, but the interference between $\a980$ and NR$_{0^+}$ is very large if the Flatt\'{e} model is replaced by the FSI model. However, if we remove NR$_{0^+}$ and refit the data, there is an obvious discrepancy between data and fit. Details can be found in Supplemental Material~\cite{supple}.

In the measurement of the absolute BF of $\lcp\to\lmdetapi$, the selection criteria are almost the same as those used to select the PWA sample except for the requirements of BDTG scores. The requirements of BDTG scores are optimized to be greater than 0.93 and 0.94 for the $\eta\to\gamma\gamma$ and $\eta\to\pip\pim\piz$ channels, respectively, by using an alternative FOM, $S/\sqrt{S+B}$. Extended unbinned maximum likelihood fits are performed to the $\mbc$ distribution, simultaneously at each energy point. In the fit, four components are considered, including signal, mismatched background, $\Lambda_{c}^+$ decay backgrounds\ which are derived from MC simulation, and combinatorial background modeled with an ARGUS function~\cite{ARGUS:1990hfq}. A truth-match method~\cite{BESIII:2023sdr} is employed to separate signals and mismatched backgrounds. The yield ratios of signals and mismatched backgrounds and $\Lambda_{c}^+$ decay backgrounds are fixed according to MC simulation. The total signal yield is given by $N_{\mathrm{sig}} = 2 \times N_{\lcp\lcm} \times \mathcal{B} \times \mathcal{B}_{\mathrm{inter}} \times \varepsilon$. Here, $N_{\lcp\lcm}$ is the number of $\lcp\lcm$ pairs calculated from the luminosities and cross sections~\cite{BESIII:2022dxl,BESIII:2022ulv,BESIII:2017kqg,BESIII:2023rwv}, and $\mathcal{B}$ is the BF of the signal decay shared for all c.m.~energy points, $\mathcal{B}_{\mathrm{inter}} = \mathcal{B}(\Lambda\to p\pim)\cdot\mathcal{B}(\eta\to\gamma\gamma)$ and $\mathcal{B}(\Lambda\to p\pim)\cdot\mathcal{B}(\eta\to\pip\pim\piz)\cdot\mathcal{B}(\piz\to\gamma\gamma)$ is the BF of intermediate decays quoted from the PDG~\cite{pdg2023}. Finally, $\varepsilon$ is the average detection efficiency based on PWA signal MC samples in which $\Lambda_c^+$ decays follow decay amplitudes with parameters fixed by PWA results: $(13.73\pm0.02)\%$ and $(4.83\pm0.01)\%$ for the $\eta\to\gamma\gamma$, and $\eta\to\pip\pim\piz$ channels, respectively. The uncertainties are statistical only. The BF is determined to be $(1.94\pm0.07)\%$ which is consistent with the previous measurements~\cite{pdg2023,BESIII:2018qyg,Belle:2020xku}. The fit plots can be found in Supplemental Material~\cite{supple}.

The systematic uncertainties on the measurement of the FFs and decay asymmetry parameters include the fixed parameters, barrier radius, additional resonant components, $\lcp$ polarization, fit method, differences between data and MC simulation, and background descriptions. The total systematic uncertainty on the BF measurement is evaluated to be 5.7\% including tracking (0.9\%), PID (0.3\%), $\Lambda$ reconstruction (2.6\%), $\eta$ reconstruction (1.0\%), BDTG score requirements (1.1\%), signal model (2.7\%), fit model (0.9\%), $\mathcal{B}_{\mathrm{inter}}$ (0.9\%), $N_{\lcp\lcm}$ (3.9\%), and MC statistics (0.4\%). Details for both the PWA and BF results can be found in Supplemental Material~\cite{supple}.

In summary, based on $6.1~\mathrm{fb}^{-1}$ of $e^+e^-$ annihilation data collected at the c.m.~energy region between 4.600 and 4.843~GeV with the BESIII detector, the first PWA of $\Lambda_c^+\to\Lambda\pi^+\eta$ is performed. The $\Lambda_c^+\to\Lambda a_0(980)^+$ decay is observed for the first time, with a statistical significance of $13.1\sigma$, and evidence for the potential pentaquark state $\Sigma(1380)^+$ is found in the $\Lambda\pi^+$ system  via a PWA, with a statistical significance larger than $3\sigma$. The BF of $\lcp\to\lmdetapi$ is measured to be $(1.94\pm0.07_{\rm stat}\pm0.01_{\rm syst})\%$, which is consistent with the previous results of BESIII~\cite{BESIII:2018qyg} and Belle~\cite{Belle:2020xku}. The product BF of $\Lambda_{c}^{+} \rightarrow \Lambda \a980$ and $\a980 \rightarrow \pi^{+}\eta$ is calculated to be $(1.05\pm0.16\pm0.05\pm0.07)\%$, where the first and second uncertainties are quoted from those of FF value, while the third is due to $\br{\lcp\to\lmdetapi}$. Taking ${\mathcal B}[a_0(980)^+\to \pi^+\eta]= 0.853\pm0.014$~\cite{bfa980}, the BF of $\Lambda_{c}^{+} \rightarrow \Lambda \a980$ is determined to be $(1.23\pm0.21)\%$, which differs significantly from the theoretical predictions evaluated in Refs.~\cite{Sharma:2009zze,Yu:2020vlt} by 1-2 orders of magnitude. A comparable scenario has been seen in $D_s^+\to a_0(980)^{+(0)}\pi^{0(+)}$ decay~\cite{BESIII:2019jjr}. Nevertheless, that puzzle can be resolved by accounting for a long-distance contribution~\cite{Hsiao:2019ait,Molina:2019udw}. However, the BF and line shape evaluated from this long-distance effect fail to adequately describe the experimental data of $\lcp\to\lambdaa$ decay. Such a large difference between theory and experiment suggests some unknown decay mechanisms. In addition, this large BF implies that $\Lambda_c^+$ decays may offer a new window to study the light scalar meson $\a980$.

Furthermore, we determine $\mathcal{B}[\Lambda_{c}^{+} \rightarrow \sgm1385 \eta]=(6.78\pm0.58\pm0.16\pm0.47)\times10^{-3}$ and $\mathcal{B}[\Lambda_{c}^{+} \rightarrow \lmd1670 \pi^{+}] \cdot \mathcal{B}[ \lmd1670 \rightarrow \Lambda \eta]=(2.74\pm0.54\pm0.24\pm0.18)\times10^{-3}$, where the third uncertainty of $\mathcal{B}[\Lambda_{c}^{+} \rightarrow \sgm1385 \eta]$ also includes the uncertainty from $\mathcal{B}[\sgm1385\to\Lambda \pip] = (87.5\pm1.5)\%$~\cite{pdg2023}. The obtained product $\mathcal{B}[\Lambda_{c}^{+} \rightarrow \sgm1385 \eta]$ is consistent with the previous BESIII result~\cite{BESIII:2018qyg} within $2\sigma$ but differs from the Belle result~\cite{Belle:2020xku} by over $3\sigma$. The obtained $\mathcal{B}[\Lambda_{c}^{+} \rightarrow \lmd1670 \pi^{+}]\cdot\mathcal{B}[\lmd1670 \rightarrow \Lambda \eta]$ is consistent with the Belle result within $1\sigma$. The $\mathcal{B}[\Lambda_{c}^{+} \rightarrow \sgm1385 \eta]$ measured in this work is in good agreement with recent calculations~\cite{Hsiao:2020iwc,Geng:2019awr}, while it differs from the early calculations~\cite{Korner:1992wi,Sharma:1996sc} by over $3\sigma$. There is a pure nonfactorizable contribution in $\Lambda^+_c\to \Sigma(1385)^+\eta$~\cite{Hsiao:2020iwc} that is difficult to calculate; our measurement is crucial to calibrate theoretical treatments of this nonfactorizable contribution. Based on the PWA results, the decay asymmetry parameters of these three intermediate processes are determined for the first time. The measured decay asymmetry parameter of $\lcp\to\sgm1385\eta$, $-0.61\pm0.15\pm0.04$, is consistent with $-0.97^{+0.43}_{-0.03}$ evaluated in Ref.~\cite{Geng:2019awr}. However, that of $\lcp\to\lambdaa$ is close to $-1$, which contradicts the small asymmetry estimated in Ref.~\cite{Sharma:2009zze}. This discrepancy might indicate issues in the consideration of $\a980$ decay constant or parity-violating transition amplitudes. Our results are essential to improve the current understanding of the dynamics of the hadronic $\lcp$ decays.


We thank En Wang, Guan-Ying Wang, Yu-Kuo Hsiao and Jia-Jun Wu for useful discussions.
The BESIII Collaboration thanks the staff of BEPCII and the IHEP computing center for their strong support. This work is supported in part by National Key R\&D Program of China under Contracts Nos. 2020YFA0406400, 2020YFA0406300, 2023YFA1606000; National Natural Science Foundation of China (NSFC) under Contracts Nos. 11635010, 11735014, 11835012, 11935015, 11935016, 11935018, 11961141012, 12025502, 12035009, 12035013, 12061131003, 12192260, 12192261, 12192262, 12192263, 12192264, 12192265, 12221005, 12225509, 12235017; the Chinese Academy of Sciences (CAS) Large-Scale Scientific Facility Program; the CAS Center for Excellence in Particle Physics (CCEPP); Joint Large-Scale Scientific Facility Funds of the NSFC and CAS under Contract No. U1832207; CAS Key Research Program of Frontier Sciences under Contracts Nos. QYZDJ-SSW-SLH003, QYZDJ-SSW-SLH040; 100 Talents Program of CAS; The Institute of Nuclear and Particle Physics (INPAC) and Shanghai Key Laboratory for Particle Physics and Cosmology; European Union's Horizon 2020 research and innovation programme under Marie Sklodowska-Curie grant agreement under Contract No. 894790; German Research Foundation DFG under Contracts Nos. 455635585, Collaborative Research Center CRC 1044, FOR5327, GRK 2149; Istituto Nazionale di Fisica Nucleare, Italy; Ministry of Development of Turkey under Contract No. DPT2006K-120470; National Research Foundation of Korea under Contract No. NRF-2022R1A2C1092335; National Science and Technology fund of Mongolia; National Science Research and Innovation Fund (NSRF) via the Program Management Unit for Human Resources \& Institutional Development, Research and Innovation of Thailand under Contract No. B16F640076; Polish National Science Centre under Contract No. 2019/35/O/ST2/02907; The Swedish Research Council; U. S. Department of Energy under Contract No. DE-FG02-05ER41374.





\onecolumngrid

\clearpage
\newpage
\appendix

\textbf{\boldmath\large Supplemental Material for ``Observation of $\Lambda_c^+ \to \Lambda a_0(980)^+$ and Evidence for $\Sigma(1380)^+$ in $\Lambda_c^+ \to \Lambda \pi^+ \eta$"}

\begin{appendices}

\section{Fit results for different PWA samples}

\begin{figure}[htbp]
	\centering
    \includegraphics[width=0.24\textwidth]{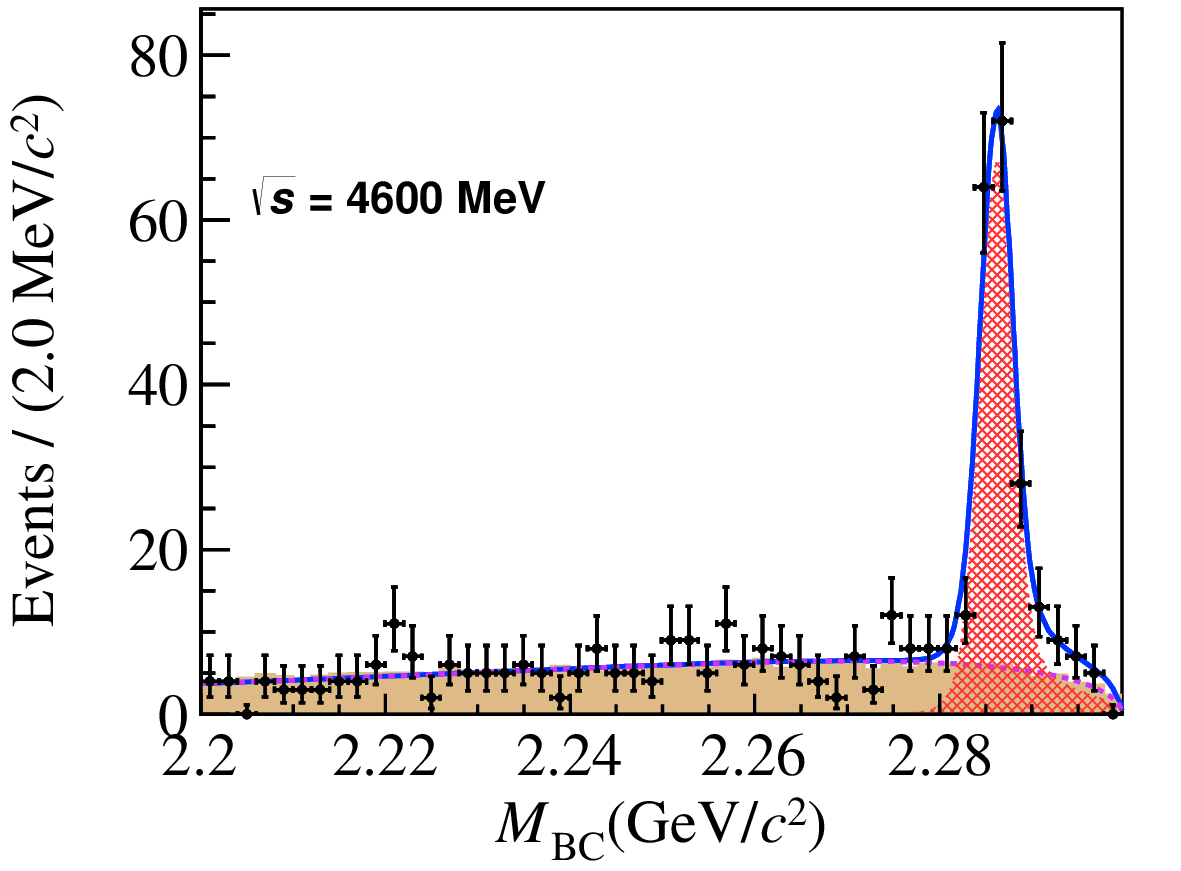}
    \includegraphics[width=0.24\textwidth]{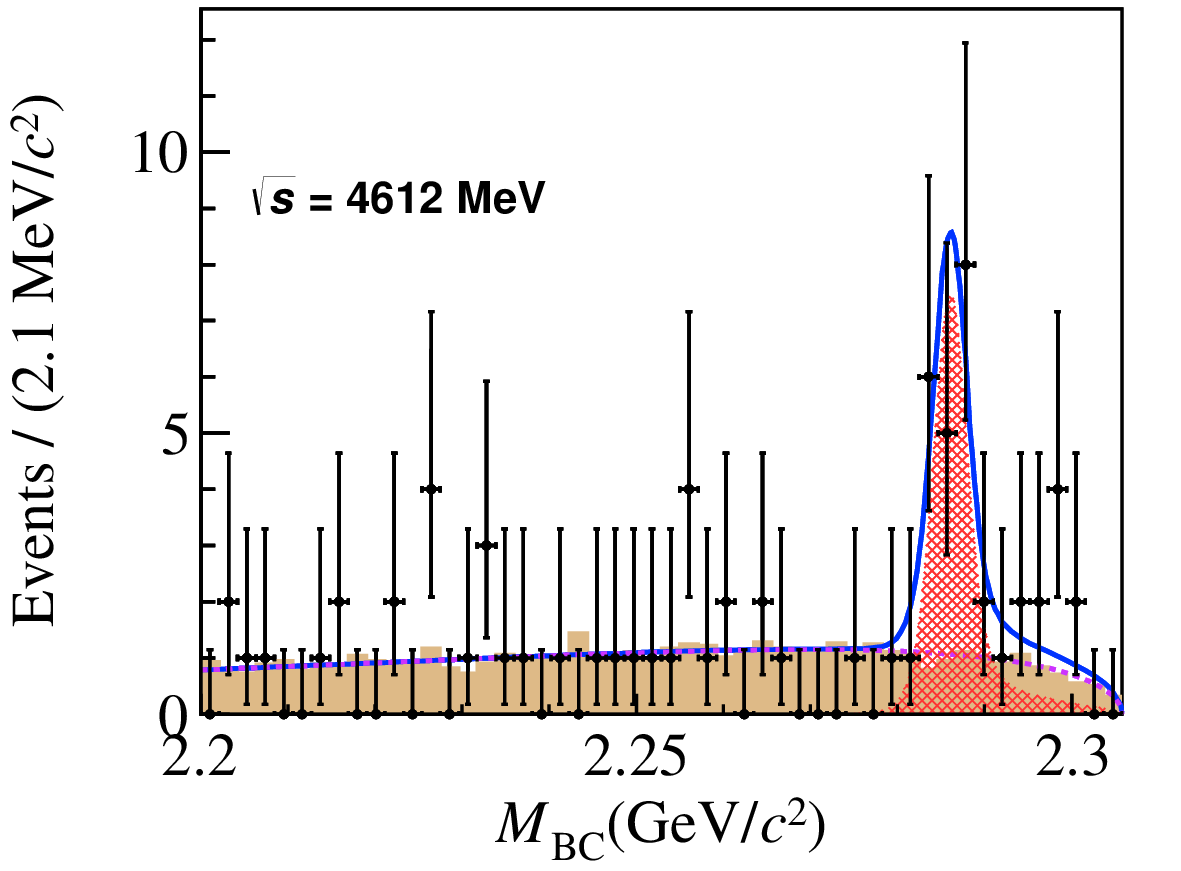}
    \includegraphics[width=0.24\textwidth]{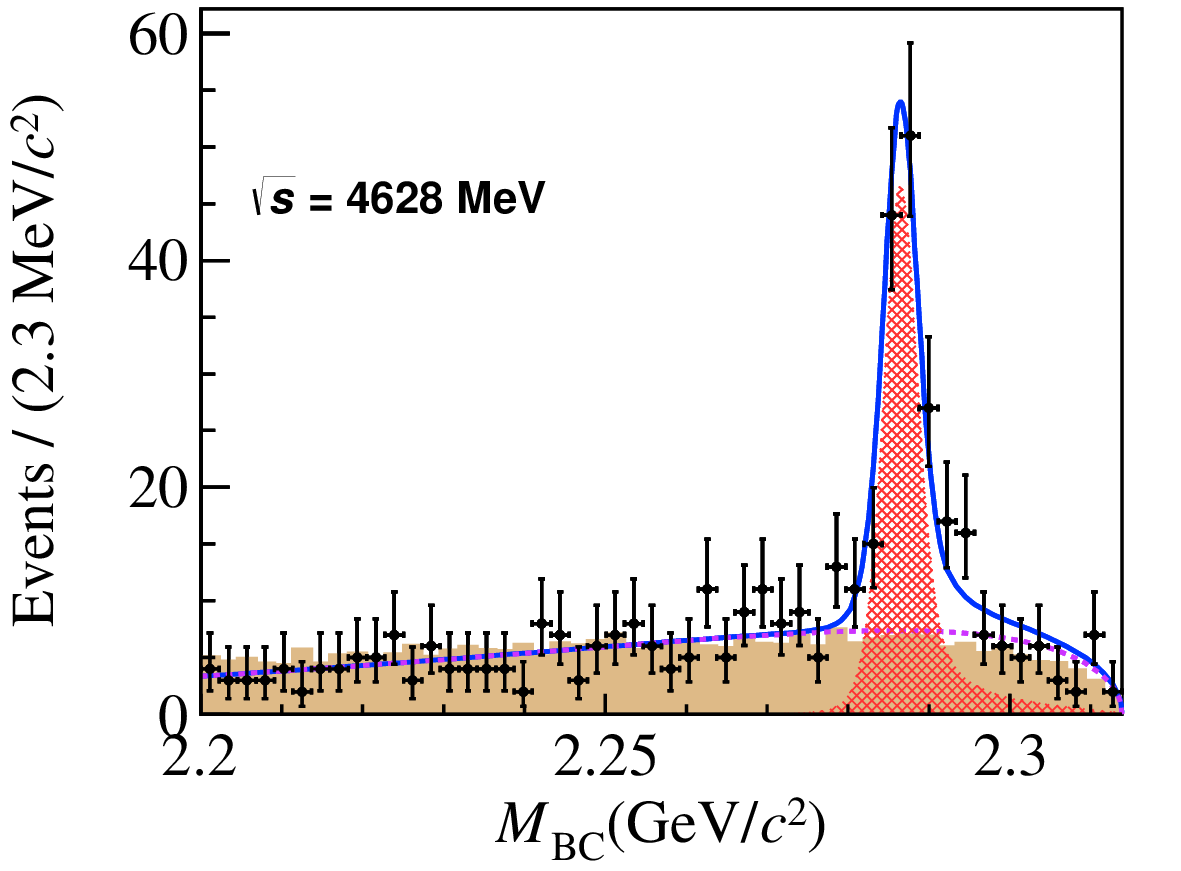}
    \includegraphics[width=0.24\textwidth]{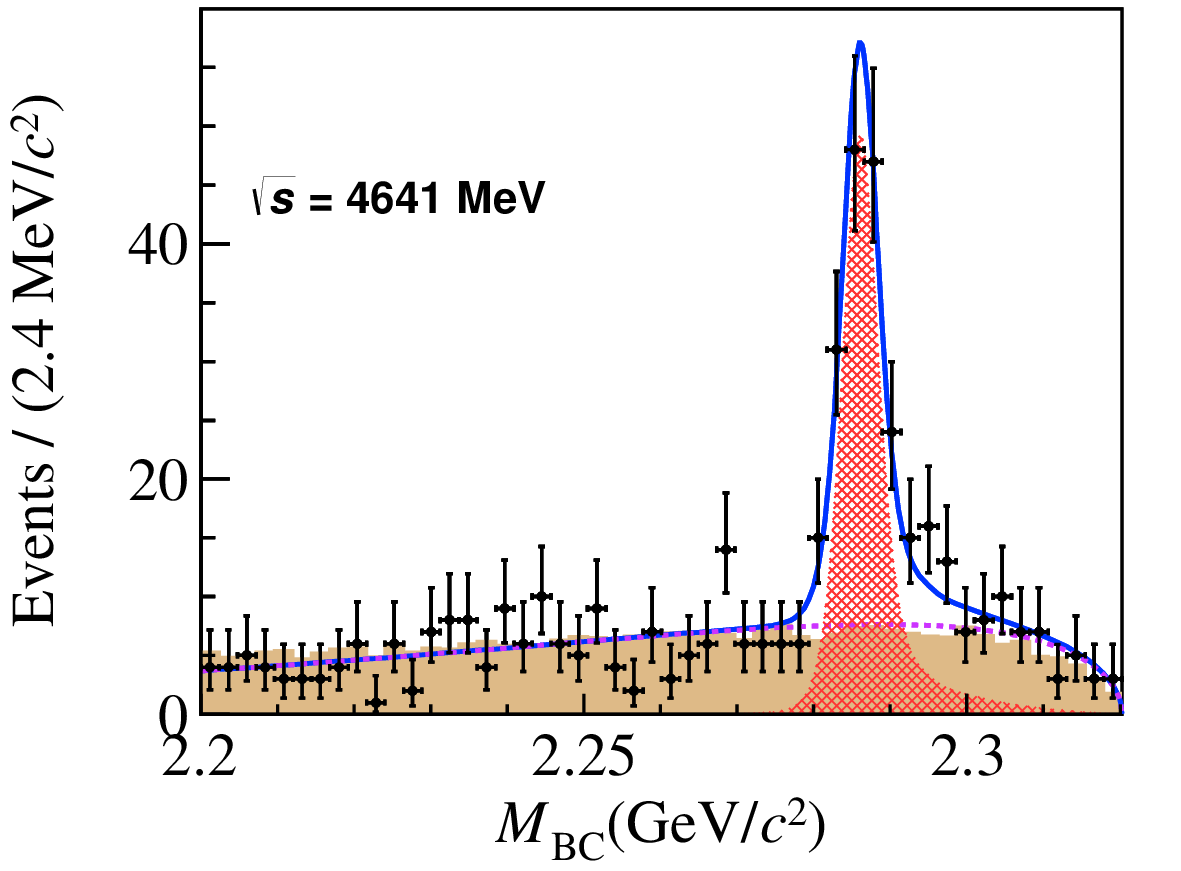}
    \includegraphics[width=0.24\textwidth]{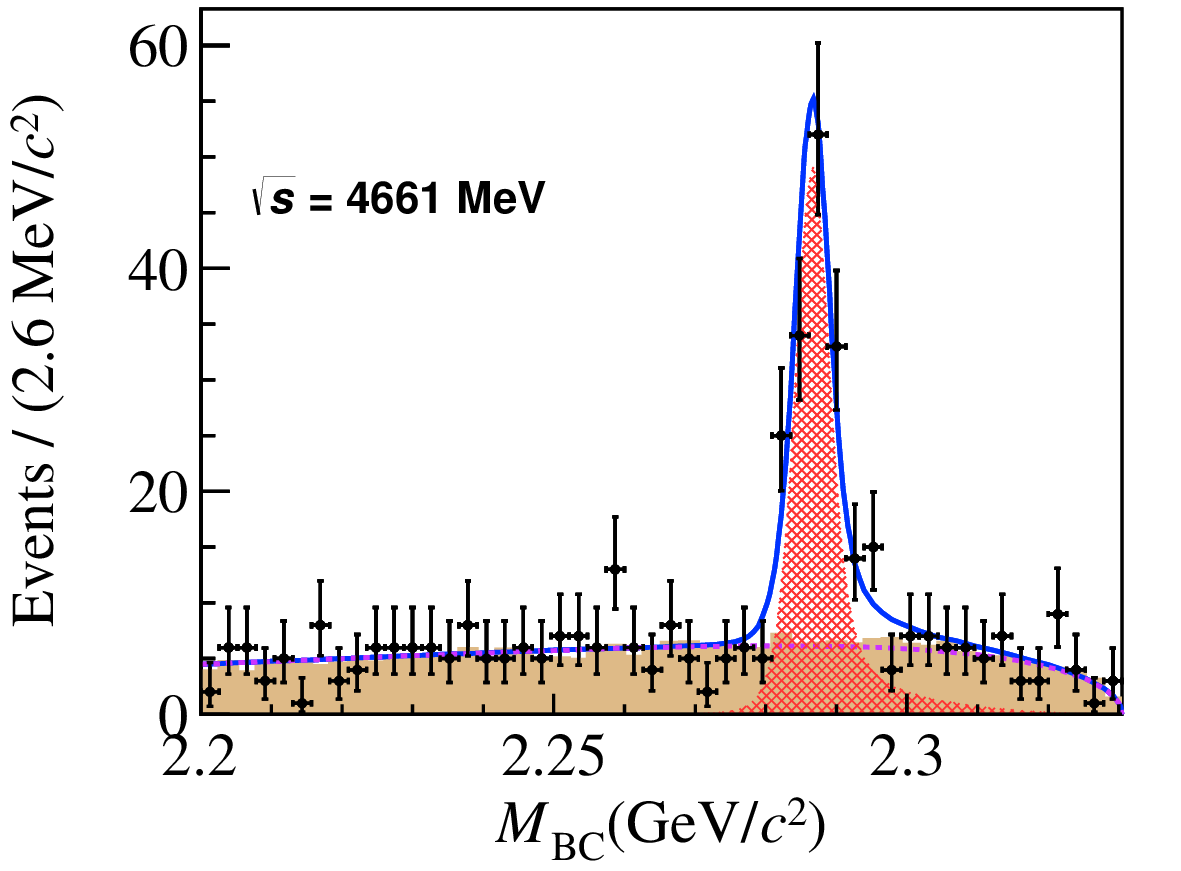}
    \includegraphics[width=0.24\textwidth]{Figure/mbc_fit/fit_mBC_4680.eps}
    \includegraphics[width=0.24\textwidth]{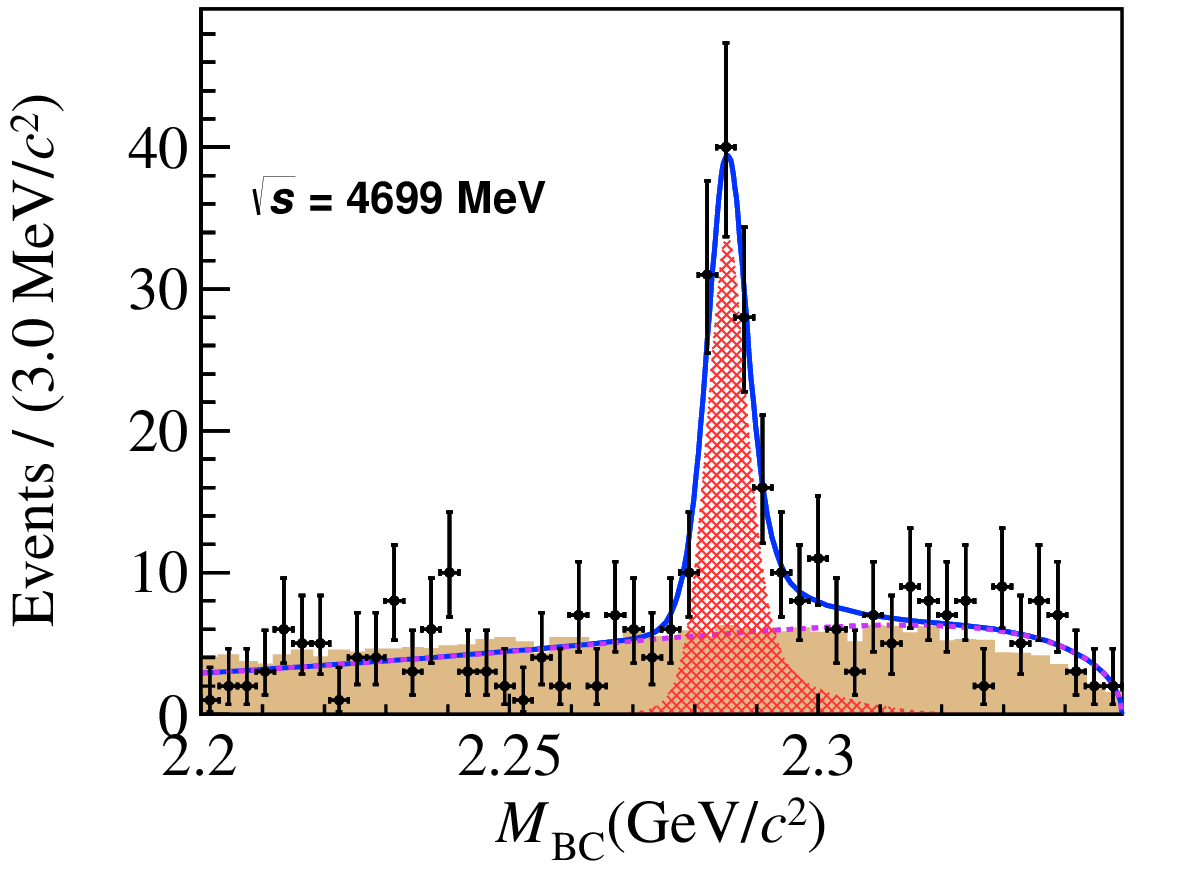}
    \includegraphics[width=0.24\textwidth]{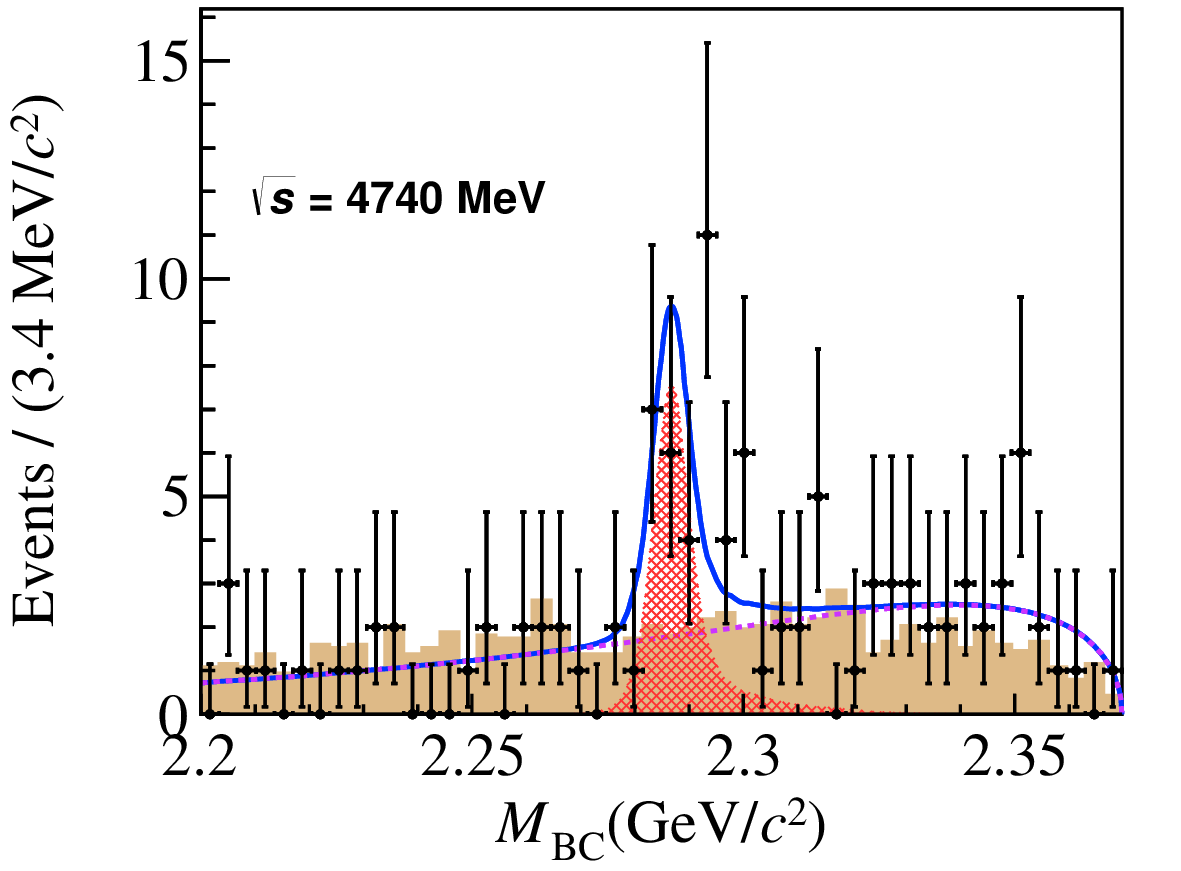}
    \includegraphics[width=0.24\textwidth]{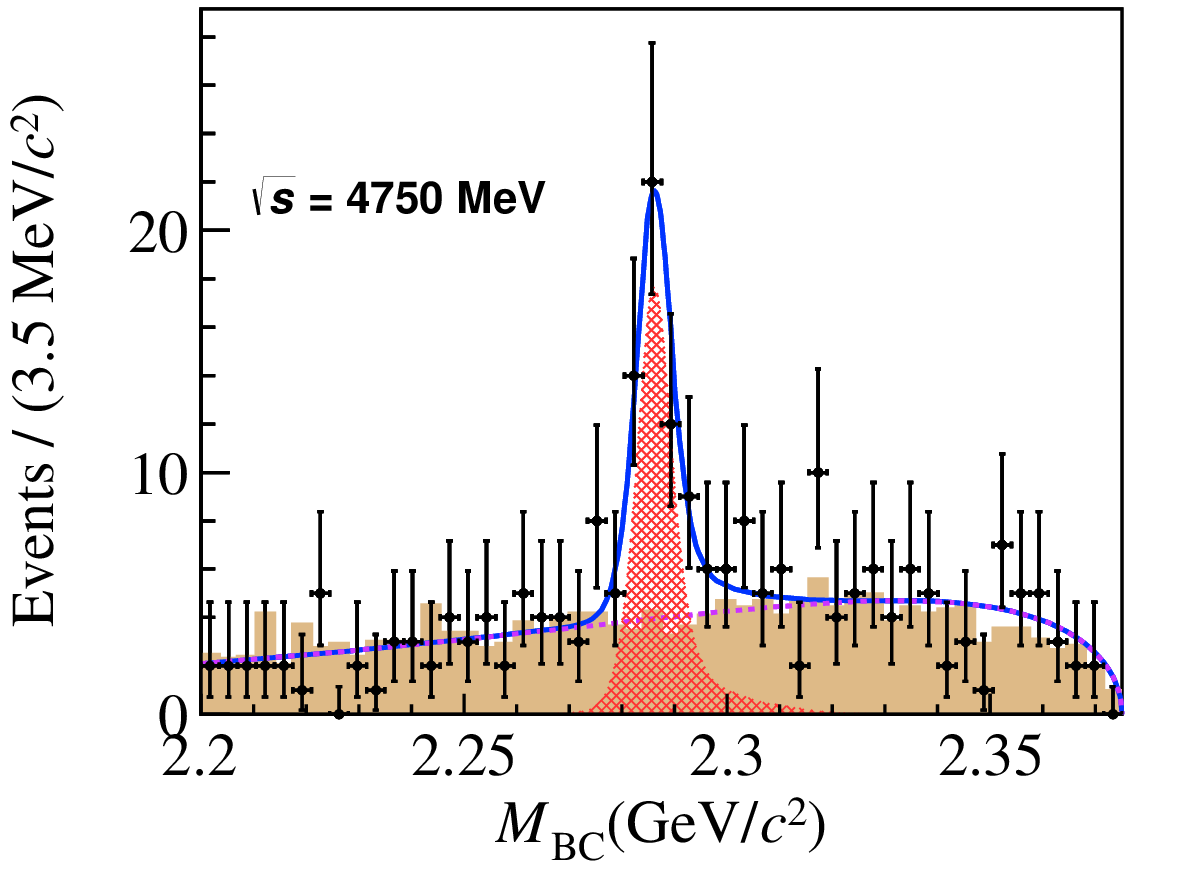}
    \includegraphics[width=0.24\textwidth]{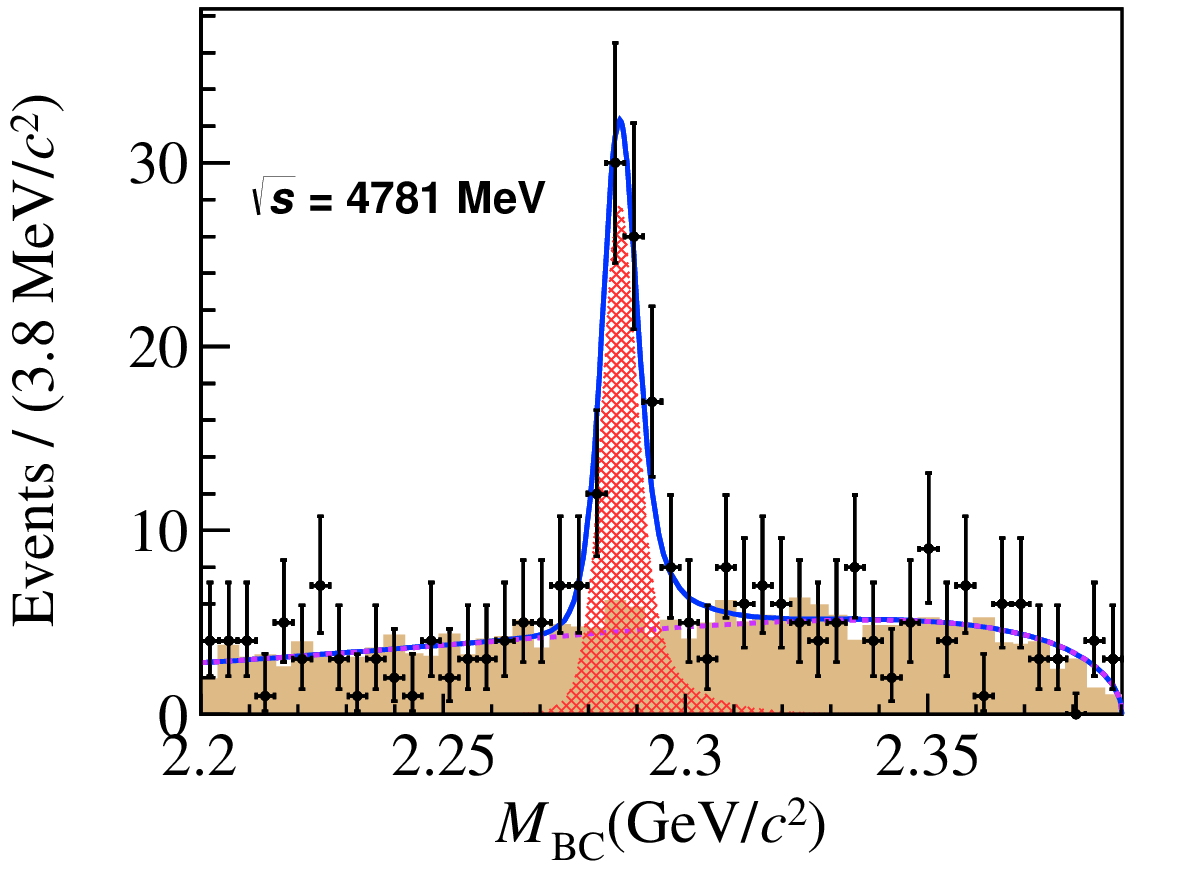}
    \includegraphics[width=0.24\textwidth]{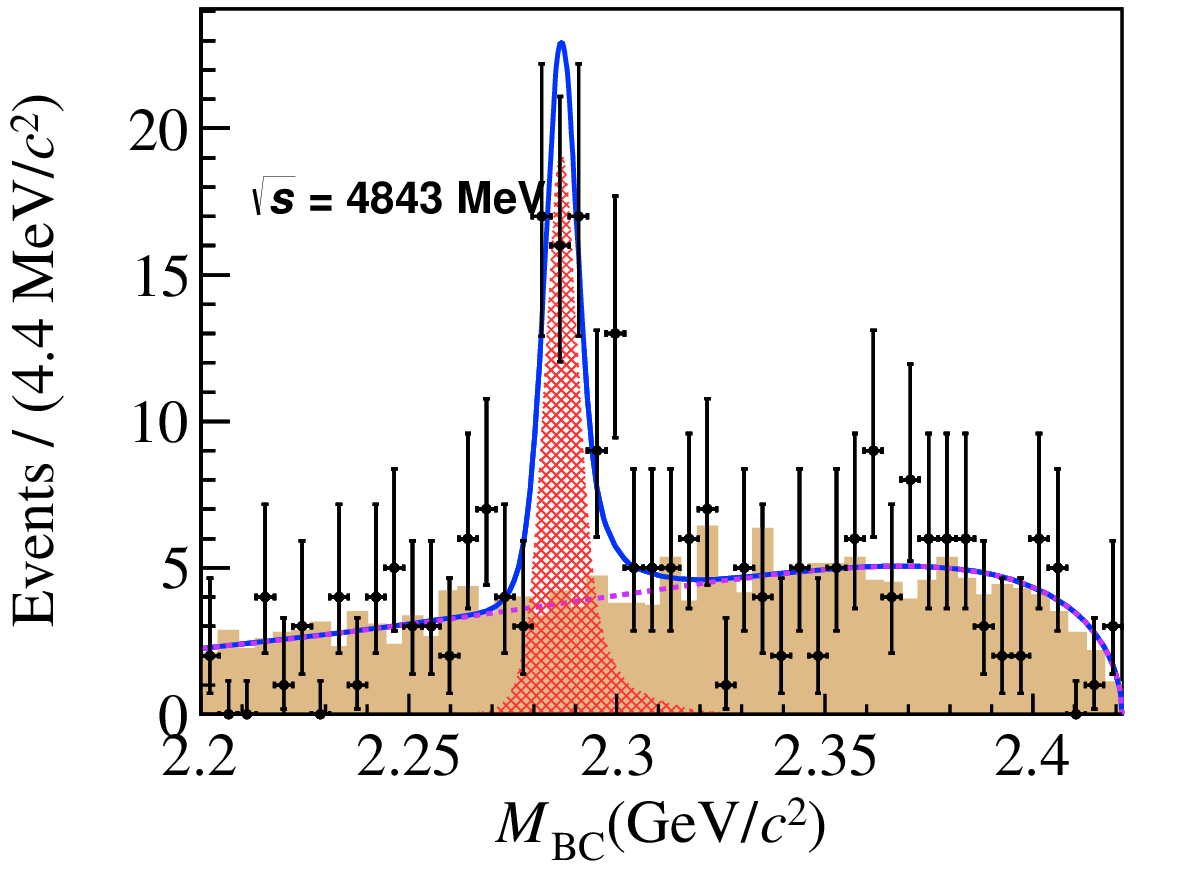}
	\caption{The fits to the $M_{\rm BC}$ distributions for $\eta\to\gamma\gamma$ and $\eta\to\pi^+\pi^-\pi^0$ channels combined from the different energy points. The points with error bars are data, the brown solid histograms are MC-simulated background, the red hatched histograms are signal, the violet dashed lines are background shapes, and the blue lines show the total fit. \label{fig:MbcfitsEcm}}
\end{figure}

Table \ref{tab:BDTG_mbcfit} summarizes the results of the $\mbc$ fits to 
the data from different c.m.~energies.  The individual fits are shown in 
Fig. \ref{fig:MbcfitsEcm}. 

\begin{table*}[htbp]
	\centering
	\caption{The results of fits to $M_{\mathrm{BC}}$ distributions from different energy points, along with $M_{\mathrm{BC}}$ signal regions, signal yields, and signal purities for the $\eta\to\gamma\gamma$ and $\eta\to\pip\pim\piz$ channels. The uncertainties are statistical only.\label{tab:BDTG_mbcfit}}
	\normalsize
	$\begin{array}{c|cr@{\,\pm\,}lr@{\,\pm\,}l|r@{\,\pm\,}lr@{\,\pm\,}l}
	\hline \hline 
    \multirow{2}{*}{\text {Sample~(MeV)}} & \multirow{2}{*}{\textit{$M_{\mathrm{BC}}~ (\mathrm{GeV}/c^2)$}} & \multicolumn{4}{c}{\textit{$\eta\to\gamma\gamma$}} & \multicolumn{4}{c}{\textit{$\eta\to\pip\pim\piz$}} \\
    \cline{3-10}
    & & \multicolumn{2}{c}{\text{Yields}} & \multicolumn{2}{c}{\text{Purities}~(\%)} & \multicolumn{2}{c}{\text{Yields}} & \multicolumn{2}{c}{\text{Purities}~(\%)} \\
	\hline 
	4600 & (2.282,2.291) & 131 & 12 & 86.2 & 1.5 & 27 & 6 & 85.6 & 3.9\\
	4612 & (2.282,2.291) & 15 & 5   & 77.8 & 6.8 & 3 & 1  & 93.5 & 4.3\\
	4628 & (2.282,2.291) & 89 & 11  & 77.1 & 2.8 & 21 & 5 & 88.8 & 3.7\\
	4641 & (2.282,2.292) & 95 & 11  & 77.2 & 2.8 & 26 & 6 & 87.8 & 3.5\\
	4661 & (2.282,2.292) & 92 & 11  & 82.0 & 2.2 & 24 & 5 & 87.9 & 3.6\\
	4682 & (2.281,2.293) & 265 & 19 & 77.0 & 1.6 & 50 & 8 & 81.8 & 3.2\\
	4700 & (2.279,2.293) & 78 & 10  & 77.2 & 3.0 & 16 & 5 & 80.8 & 5.8\\
	4740 & (2.281,2.296) & 17 & 5   & 73.5 & 6.9 & 2 & 2  & 49.1 & 29.8\\
	4750 & (2.281,2.296) & 33 & 7   & 67.6 & 5.4 & 10 & 3 & 87.4 & 5.2\\
	4781 & (2.279,2.297) & 60 & 9   & 76.8 & 3.3 & 10 & 4 & 74.7 & 8.2\\
	4843 & (2.279,2.297) & 37 & 7   & 74.5 & 4.4 & 7 & 3  & 72.4 & 10.1\\
	\hline
	\hline
	\end{array}$
\end{table*}

\section{Helicity amplitude definition}

In helicity amplitude formalism, three-body decay is described as a two-step sequential quasi-two-body decay based on isobar model. For each two-body decay $0\to1+2$, the helicity amplitude can be written as $A^{0\to1+2}_{\lambda_0,\lambda_1,\lambda_2} = H^{0\to1+2}_{\lambda_1,\lambda_2}D^{J_0*}_{\lambda_0,\lambda_1-\lambda_2}(\phi,\theta,0)$, where the helicity coupling $H^{0\to1+2}_{\lambda_1,\lambda_2}$ is given by the LS coupling formula~\cite{suppleChung:1993da,suppleChung:1997jn,suppleChung:2003fk} along with barrier factor terms
\begin{equation}
H^{0\to1+2}_{\lambda_1,\lambda_2} = \sum_{ls} g_{ls} \sqrt{\frac{2l+1}{2J_{0}+1}} \bra{l0,s\delta}\ket{J_{0},\delta} \times\bra{J_1J_2,\lambda_1-\lambda_2}\ket{s,\delta}  (\frac{q}{q_0})^{l}B_{l}^{\prime}(q,q_0,d),
\label{equ:LS}
\end{equation}
and $D^{J_{0}*}_{\lambda_0,\lambda_1-\lambda_2}$ is the Wigner $D$-function, $\phi$ and $\theta$ are helicity angles. The $g_{ls}$ is the partial wave amplitude, $J_{0,1,2}$ are the spins of the particles 0, 1, and 2, $\lambda_{1,2}$ are the helicities for the particles 1 and 2, and $\delta=\lambda_1-\lambda_2$ is the helicity difference. Here, $q$ is the three-momentum modulus of particle 1 in the rest frame of particle 0, which is calculated as $q=\frac{\sqrt{\left[m^2-\left(m_1+m_2\right)^2\right]\left[m^2-\left(m_1-m_2\right)^2\right]}}{2 m}$, where $m$, $m_1$ and $m_2$ are the masses of the particles 0, 1, and 2, respectively. The normalization factor $q_0$ is calculated at the nominal resonance mass. The factor $B_{l}^{\prime}(q,q_0,d)$ is the reduced Blatt-Weisskopf barrier factor~\cite{suppleVonHippel:1972fg}, which is explicitly expressed as
\begin{equation}
\begin{aligned}
& B_0^{\prime}\left(q, q_0, d\right)=1, \\
& B_1^{\prime}\left(q, q_0, d\right)=\sqrt{\frac{1+\left(q_0 d\right)^2}{1+(q d)^2}}, \\
& B_2^{\prime}\left(q, q_0, d\right)=\sqrt{\frac{9+3\left(q_0 d\right)^2+\left(q_0 d\right)^4}{9+3(q d)^2+(q d)^4}}, \\
&
\label{equ:barrier_factor}
\end{aligned}
\end{equation}
and the radius of the centrifugal barrier $d$ is chosen as $d=0.73~\mathrm{fm}$~\cite{suppleBESIII:2022udq}. For decay chains with resonant intermediate states, the barrier factor term is included. For those with NR intermediate states, the barrier factor term is omitted.

The $\lcp\to\lambdaa,\a980\to\pip\eta,\Lambda\to p\pim$ decay amplitude is constructed as
\begin{equation}
    A_{\lambda_{\Lambda_c^{+}}, \lambda_p}^{a_0^+}=\sum_{\lambda_{\Lambda}} A_{\lambda_{\Lambda_c^{+}}, 0, \lambda_{\Lambda}}^{\Lambda_{c}^{+} \rightarrow a_0^+ \Lambda} R_{a_0^+}\left(M_{\pi^{+} \eta}\right) A_{0, 0,0}^{a_0^+ \rightarrow \pi^{+} \eta} A_{\lambda_{\Lambda}, \lambda_p, 0}^{\Lambda \rightarrow p \pi^{-}},
\end{equation}
where $R_{a_0^+}$ is the propagator of the $\a980$ described by the two coupled-channel Flatt\'{e} model~\cite{suppleFlatte:1976xu},
\begin{equation}
    R_{a_0^+}(m) = \frac{1}{m_0^2-m^2-i(g_1\rho_{\eta\pi}(m)+g_2\rho_{K\bar{K}}(m))}.
\end{equation}
The nominal mass $m_0$ and coupling constants of the $\a980$ decaying to the $\eta\pi$, $g_1$, and $K\bar{K}$, $g_2$, coupled channels are quoted from Ref.~\cite{suppleBESIII:2016tqo}. For the NR decay, the dynamical function is set to be unity, so the decay amplitude is
\begin{equation}
    A_{\lambda_{\Lambda_c^{+}}, \lambda_p}^{N\!R}=\sum_{\lambda_{\Lambda}} A_{\lambda_{\Lambda_c^{+}}, 0, \lambda_{\Lambda}}^{\Lambda_{c}^{+} \rightarrow N\!R \Lambda} \left(M_{\pi^{+} \eta}\right) A_{0, 0,0}^{N\!R \rightarrow \pi^{+} \eta} A_{\lambda_{\Lambda}, \lambda_p, 0}^{\Lambda \rightarrow p \pi^{-}}.
\end{equation}
For the decay through the $\Sigma^{*+}$ and $\Lambda^{*}$ intermediate states, the amplitudes are written as
\begin{equation}
    A^{\Sigma^{*+}}_{\lambda_{\lcp},\lambda_p} = \sum_{\lambda_{\Sigma^{*+}},\lambda_{\Lambda}}A_{\lambda_{\Lambda_c^{+}}, \lambda_{\Sigma^{*+}}, 0}^{\Lambda_{c}^{+} \rightarrow \Sigma^{*+}\eta} 
    R_{\Sigma^{*+}}(M_{\Lambda\pip})
    A_{\lambda_{\Sigma^{*+}}, \lambda_{\Lambda},0}^{\Sigma^{*+} \rightarrow \Lambda \pi^{+}} A_{\lambda_{\Lambda}, \lambda_p, 0}^{\Lambda \rightarrow p \pi^{-}},
\end{equation}
\begin{equation}
    A^{\Lambda^{*}}_{\lambda_{\lcp},\lambda_p} = \sum_{\lambda_{\Lambda^{*}},\lambda_{\Lambda}}A_{\lambda_{\Lambda_c^{+}}, \lambda_{\Lambda^{*}}, 0}^{\Lambda_{c}^{+} \rightarrow \Lambda^{*}\pip} 
    R_{\Lambda^{*}}(M_{\Lambda\eta})
    A_{\lambda_{\Lambda^{*}}, \lambda_{\Lambda},0}^{\Lambda^{*} \rightarrow \Lambda \eta} A_{\lambda_{\Lambda}, \lambda_p, 0}^{\Lambda \rightarrow p \pi^{-}}.
\end{equation}
Here, the relativistic Breit-Wigner (RBW) formula~\cite{suppleBESIII:2022udq} is used as the propagator,
\begin{equation}
    R_{\Sigma^{*+}/\Lambda^*} = \frac{1}{m_0^2-m^2-im_0\Gamma(m)}.
\end{equation}
The nominal mass and width of the $\sgm1385$ are fixed to the corresponding values from the PDG~\cite{supplepdg2023}, and those of $\lmd1670$ are taken from a recent measurement~\cite{suppleSarantsev:2019xxm}. The amplitude of $\Lambda \to p\pim$ is constrained according to
\begin{equation}
    \alpha_{\Lambda} =\frac{\left|H_{\frac{1}{2}, 0}^{\Lambda\to p\pi^-}\right|^2-\left|H_{-\frac{1}{2}, 0}^{\Lambda\to p\pi^-}\right|^2}{\left|H_{\frac{1}{2}, 0}^{\Lambda\to p\pi^-}\right|^2+\left|H_{-\frac{1}{2}, 0}^{\Lambda\to p\pi^-}\right|^2}=-\frac{2 \Re\left(g_{0, \frac{1}{2}}^{\Lambda\to p\pi^-} \, g_{1, \frac{1}{2}}^{\Lambda\to p\pi^-*}\right)}{\left|g_{0, \frac{1}{2}}^{\Lambda\to p\pi^-}\right|^2+\left|g_{1, \frac{1}{2}}^{\Lambda\to p\pi^-}\right|^2},
\end{equation}
where $\alpha_\Lambda$ is the decay asymmetry taken from the PDG~\cite{supplepdg2023},
The full amplitude of $\lcp$ decay is the coherent sum of amplitudes of all decay chains, and the alignment $D$-functions are considered to align the helicities of the final state protons~\cite{suppleBESIII:2018qyg,suppleWang:2020giv},
\begin{equation}
    \mathcal{A}_{\lambda_{\Lambda_c^+},\lambda_p} = A_{\lambda_{\Lambda_c^+},\lambda_p}^{a_0^+} + A_{\lambda_{\Lambda_c^+},\lambda_p}^{N\!R}+
    \sum_{\lambda_p^\prime}A_{\lambda_{\Lambda_c^+},\lambda_p}^{\Sigma^{*+}}D_{\lambda_{p}^{\prime},\lambda_p}(\alpha_p,\beta_p,\gamma_p)+
    \sum_{\lambda_p^\prime}A_{\lambda_{\Lambda_c^+},\lambda_p}^{\Lambda^{*}}D_{\lambda_{p}^{\prime},\lambda_p}(\alpha_p^{\prime},\beta_p^{\prime},\gamma_p^{\prime}).
\end{equation}

The total amplitude for polarized $\lcp$ can be written as
\begin{equation}
    \begin{aligned}
    |\mathcal{A}|^2 &=
    \sum_{\lambda_{\Lambda_c^+},\lambda_{\Lambda_c^+}^\prime,\lambda_p} \rho_{\lambda_{\Lambda_c^+},\lambda_{\Lambda_c^+}^\prime}\mathcal{A}_{\lambda_{\Lambda_c^+},\lambda_p}\mathcal{A}^{*}_{\lambda_{\Lambda_c^+},\lambda_p}\\
    &=\sum_{\lambda_p}\left[
    \left(\begin{array}{cc}
	\mathcal{A}_{-1/2,\lambda_{p}} & \mathcal{A}_{1/2,\lambda_{p}}
	\end{array}\right)
    \cdot\frac{1}{2}
    \left(\begin{array}{cc}
	1+P_z & P_x-i P_y \\
	P_x+i P_y & 1-P_z
	\end{array}\right)
    \cdot
	\left(\begin{array}{c}
	\mathcal{A}^{*}_{-1/2,\lambda_{p}} \\
	\mathcal{A}^{*}_{1/2,\lambda_{p}}
	\end{array}\right)
	\right],
    \end{aligned}
\end{equation}
where $P_x,P_y,P_z$ are the polarization components. According to Ref.~\cite{suppleChen:2019hqi}, the polarization components of $\lcp$ from $e^+e^-$ collision are fixed to $P_z=P_x=0$, $P_y(\theta_{ee},\alpha_0,\Delta_0)\propto \sqrt{1-\alpha_0^2}\sin\theta_{ee}\cos\theta_{ee}\sin\Delta_0$. Here, $\theta_{ee}$ is the polar angle of the $\lcp$ with respect to the $e^+$ beam in the $e^+e^-$ c.m.~system, $\alpha_0$ is fixed to the values from Refs.~\cite{suppleBESIII:2017kqg,suppleBESIII:2023rwv}, and $\Delta_0$ is fixed according to polarization results in data. The $\alpha_0$ and $\Delta_0$ have a relation with helicity amplitudes for $e^+e^-\to\Lambda_c^+\bar{\Lambda}_c^-$,
\begin{equation}
    \begin{aligned}
        \alpha_0 &= \frac{|H^{e^+e^-\to\lcp\lcm}_{\frac{1}{2},-\frac{1}{2}}|^2-2|H^{e^+e^-\to\lcp\lcm}_{\frac{1}{2},\frac{1}{2}}|^2}{|H^{e^+e^-\to\lcp\lcm}_{\frac{1}{2},-\frac{1}{2}}|^2+2|H^{e^+e^-\to\lcp\lcm}_{\frac{1}{2},\frac{1}{2}}|}, \\
	   \Delta_0 &= \arg(H^{e^+e^-\to\lcp\lcm}_{\frac{1}{2},-\frac{1}{2}}) - \arg(H^{e^+e^-\to\lcp\lcm}_{\frac{1}{2},\frac{1}{2}}).
	\label{equ:alph0_delta0}
    \end{aligned}
\end{equation}

\section{Definitions of decay asymmetry parameters}

The explicit definitions of the decay asymmetry parameters are given below.  

\begin{eqnarray}
    \alpha_{\Lambda\a980} &=\frac{\left|H_{\frac{1}{2}, 0}^{\lcp\to\Lambda\a980}\right|^2-\left|H_{-\frac{1}{2}, 0}^{\lcp\to\Lambda\a980}\right|^2}{\left|H_{\frac{1}{2}, 0}^{\lcp\to\Lambda\a980}\right|^2+\left|H_{-\frac{1}{2}, 0}^{\lcp\to\Lambda\a980}\right|^2}=-\frac{2 \Re\left(g_{0, \frac{1}{2}}^{\lcp\to\Lambda\a980} \, g_{1, \frac{1}{2}}^{\lcp\to\Lambda\a980*}\right)}{\left|g_{0, \frac{1}{2}}^{\lcp\to\Lambda\a980}\right|^2+\left|g_{1, \frac{1}{2}}^{\lcp\to\Lambda\a980}\right|^2}, \\
    \alpha_{\sgm1385\eta} &=\frac{\left|H_{\frac{1}{2}, 0}^{\lcp\to\sgm1385\eta}\right|^2-\left|H_{-\frac{1}{2}, 0}^{\lcp\to\sgm1385\eta}\right|^2}{\left|H_{\frac{1}{2}, 0}^{\lcp\to\sgm1385\eta}\right|^2+\left|H_{-\frac{1}{2}, 0}^{\lcp\to\sgm1385\eta}\right|^2}=-\frac{2 \Re\left(g_{1, \frac{3}{2}}^{\lcp\to\sgm1385\eta}  \, g_{2, \frac{3}{2}}^{\lcp\to\sgm1385\eta*}\right)}{\left|g_{1, \frac{3}{2}}^{\lcp\to\sgm1385\eta}\right|^2+\left|g_{2, \frac{3}{2}}^{\lcp\to\sgm1385\eta}\right|^2}, \\
    \alpha_{\lmd1670\pip} &=\frac{\left|H_{\frac{1}{2}, 0}^{\lcp\to\lmd1670\pip}\right|^2-\left|H_{-\frac{1}{2}, 0}^{\lcp\to\lmd1670\pip}\right|^2}{\left|H_{\frac{1}{2}, 0}^{\lcp\to\lmd1670\pip}\right|^2+\left|H_{-\frac{1}{2}, 0}^{\lcp\to\lmd1670\pip}\right|^2}=-\frac{2 \Re\left(g_{0, \frac{1}{2}}^{\lcp\to\lmd1670\pip} \,  g_{1, \frac{1}{2}}^{\lcp\to\lmd1670\pip*}\right)}{\left|g_{0, \frac{1}{2}}^{\lcp\to\lmd1670\pip}\right|^2+\left|g_{1, \frac{1}{2}}^{\lcp\to\lmd1670\pip}\right|^2}.
\label{equ:alpha}
\end{eqnarray}

\section{Dalitz plots for the  data and the baseline solution}

\begin{figure}[htbp]
	\centering
	\includegraphics[width=0.32\textwidth]{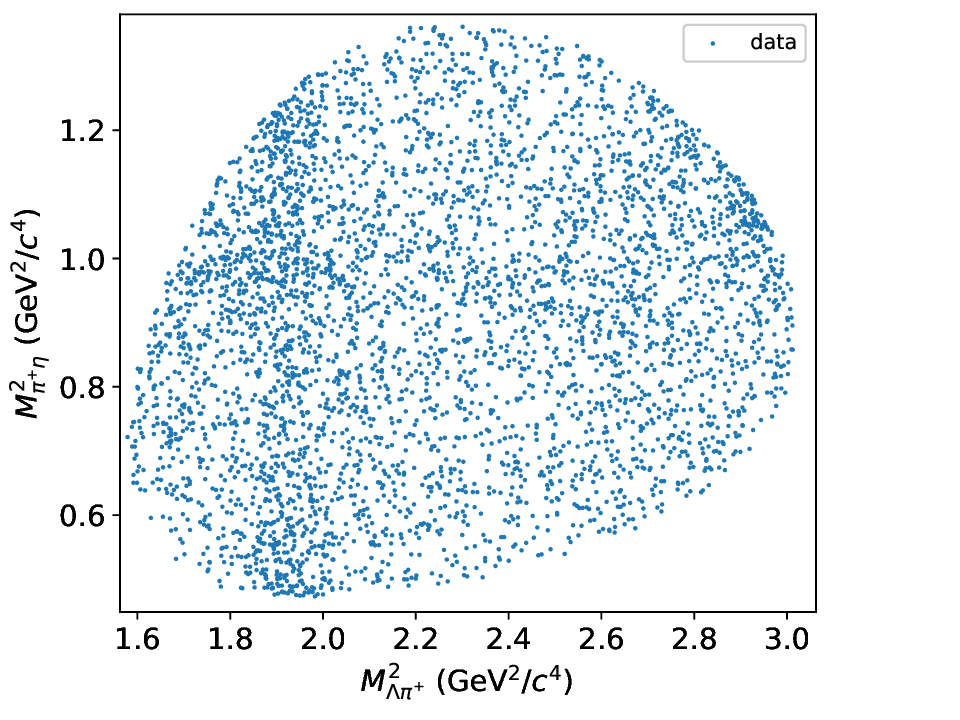}
	\includegraphics[width=0.32\textwidth]{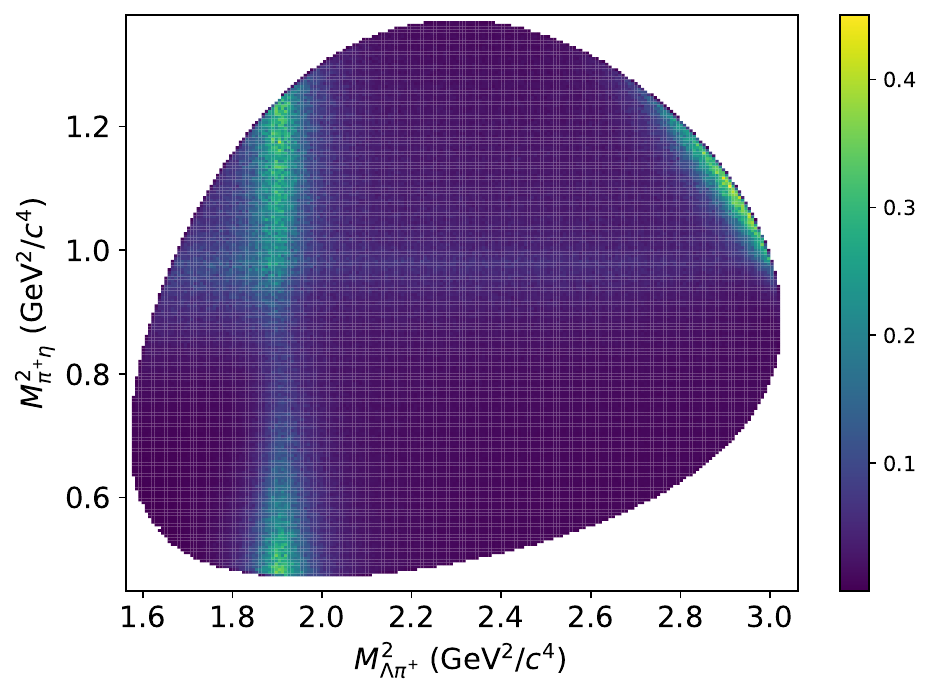}
	\caption{The Dalitz plots of data (left) and fit results (right) at all energy points.\label{fig:Dalitz}}
\end{figure}

Dalitz plots for both the  data and the baseline solution fit are 
shown in in Fig.~\ref{fig:Dalitz}. 

\section{\boldmath Fit results of PWA with $\Sigma(1380)^+$}

\begin{figure}[H]
	\centering
	\includegraphics[width=0.32\textwidth]{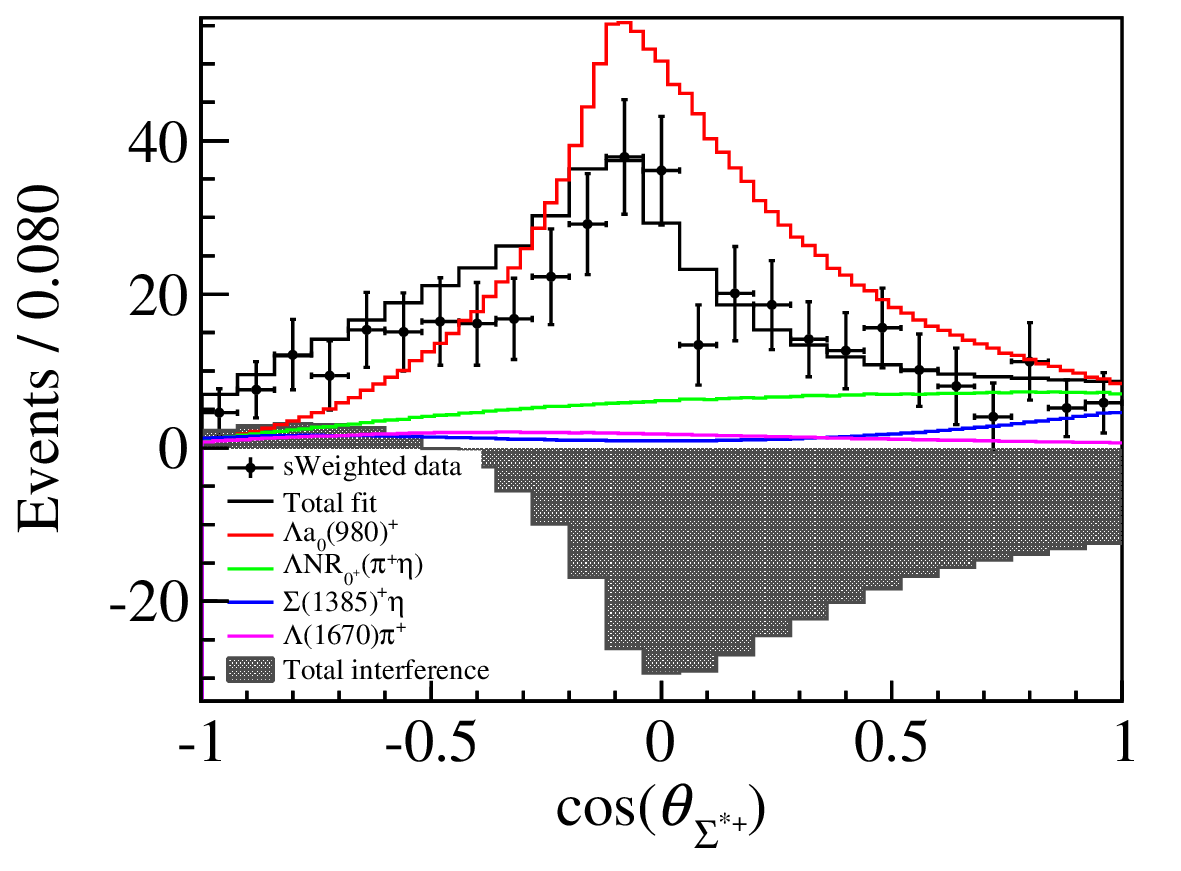}
	\includegraphics[width=0.32\textwidth]{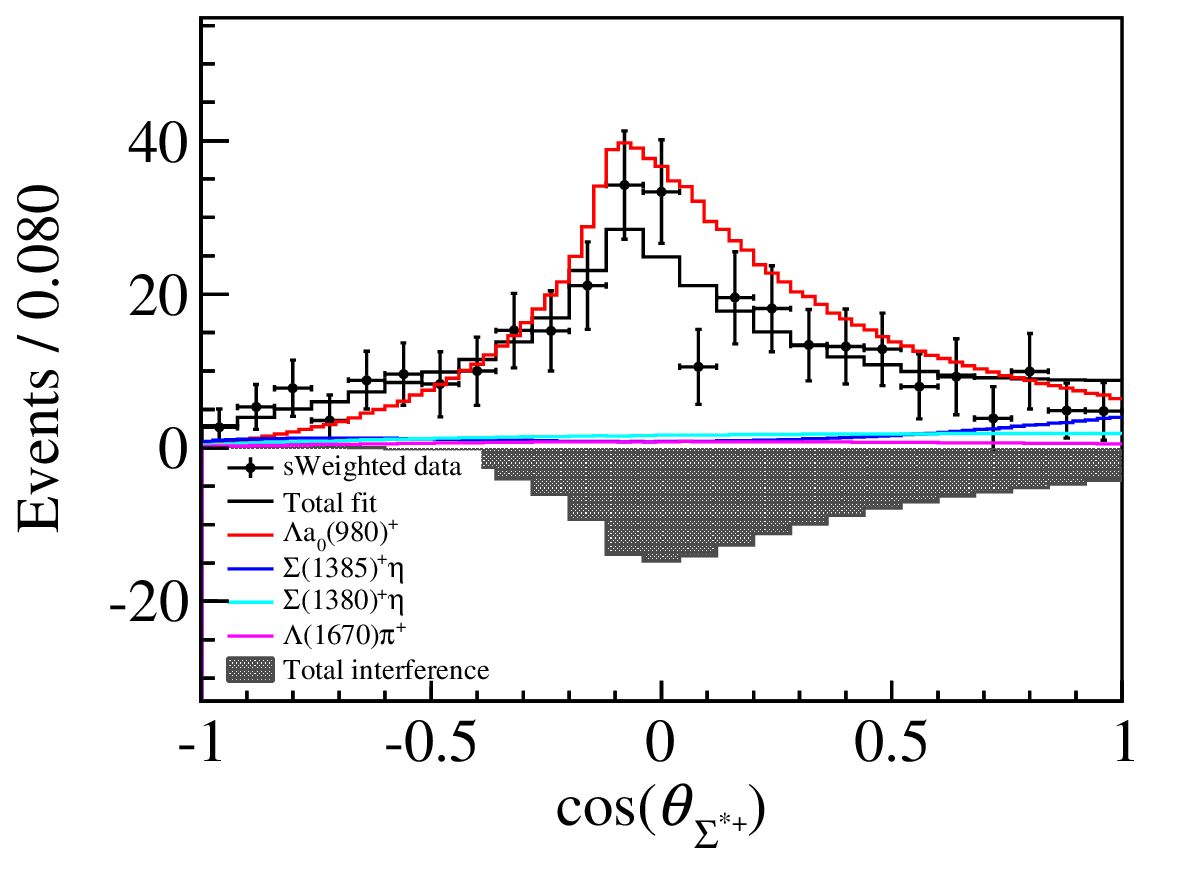}
    \includegraphics[width=0.32\textwidth]{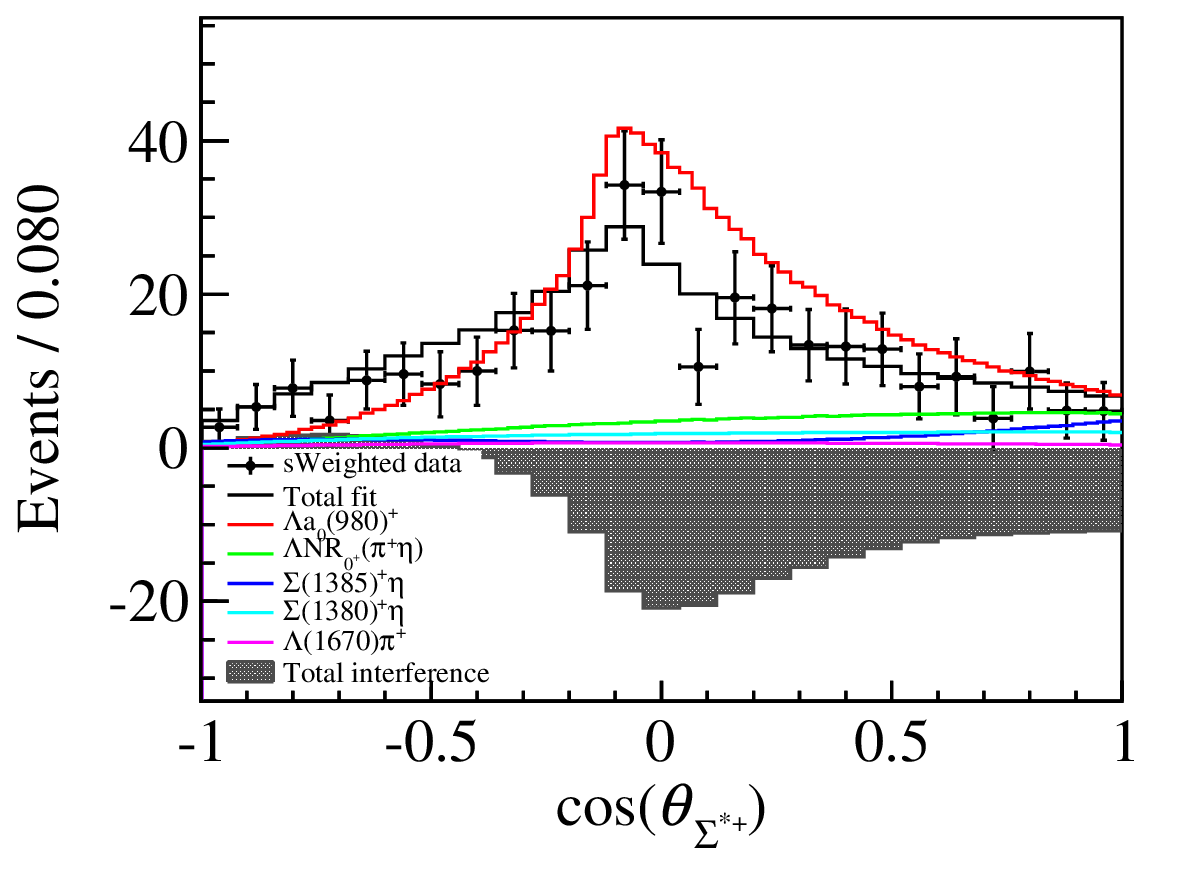}
	\caption{Projections of the fit results vs.~the $\Sigma^{*+}$ helicity angle, $\cos\theta_{\Sigma^{*+}}$ in the baseline model (left), model A (middle), and model B (right). The points with error bars are sWeighted data at all energy points. The curves in different colors are different components.\label{fig:costheta_sigmastar}}
\end{figure}

In the $\a980$ signal region in Dalitz plot, defined as $M_{\Lambda\pip}>1.44\,\gevcc$ and $M_{\Lambda\eta}>1.72\,\gevcc$, the $\Sigma^{*+}$ helicity angle, $\cos\theta_{\Sigma^{*+}}$ distribution are shown in Fig.~\ref{fig:costheta_sigmastar}. It is found that in the baseline model, the fit curve exceeds the data near $\cos\theta_{\Sigma^{*+}} \simeq -0.35$. However, in model A where the $N\!R_{0^{+}}$ amplitude is replaced by the $\Sigma(1380)^+$, the fit curve is more consistent with data.

\section{\boldmath Test $\lmd1670$ and $\a980$ line-shapes of FSI model}

\begin{figure}[htbp]
	\centering
	\includegraphics[width=0.32\textwidth]{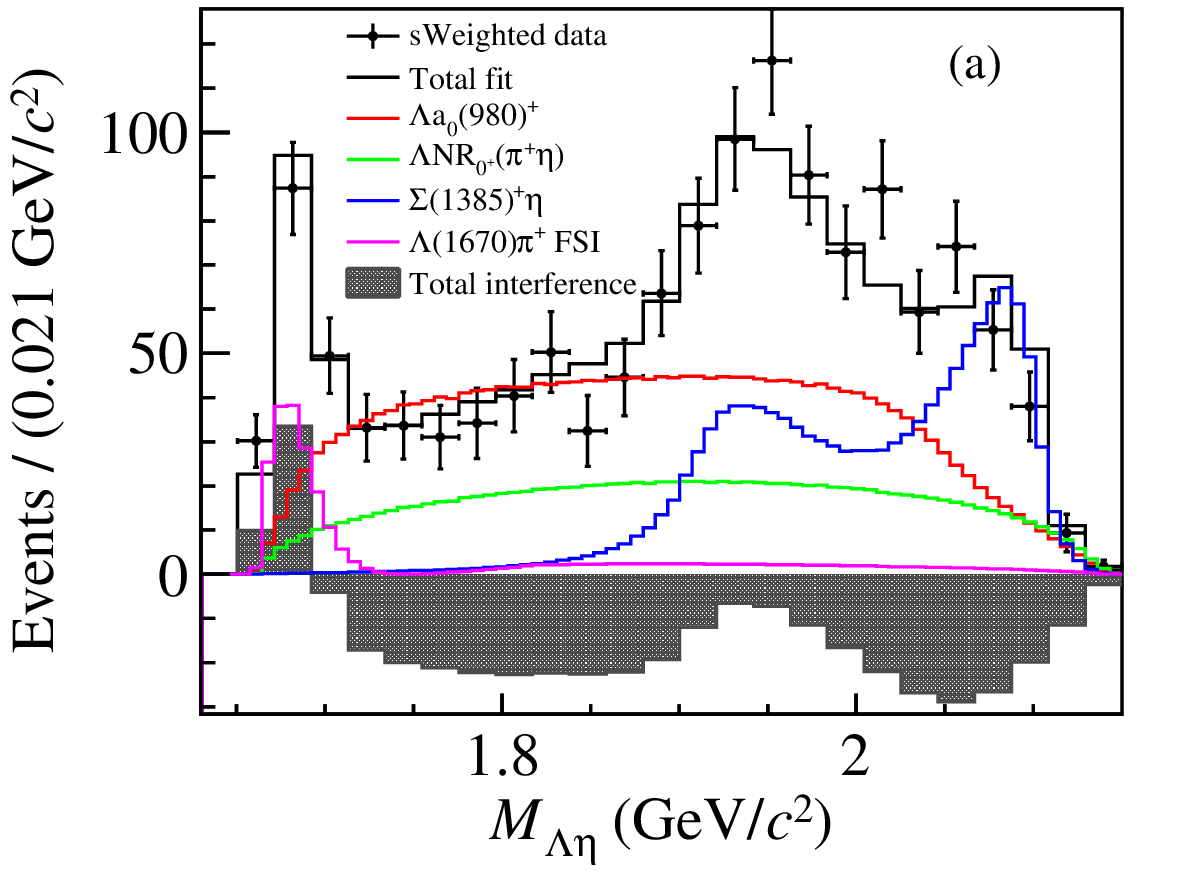}
 	\includegraphics[width=0.32\textwidth]{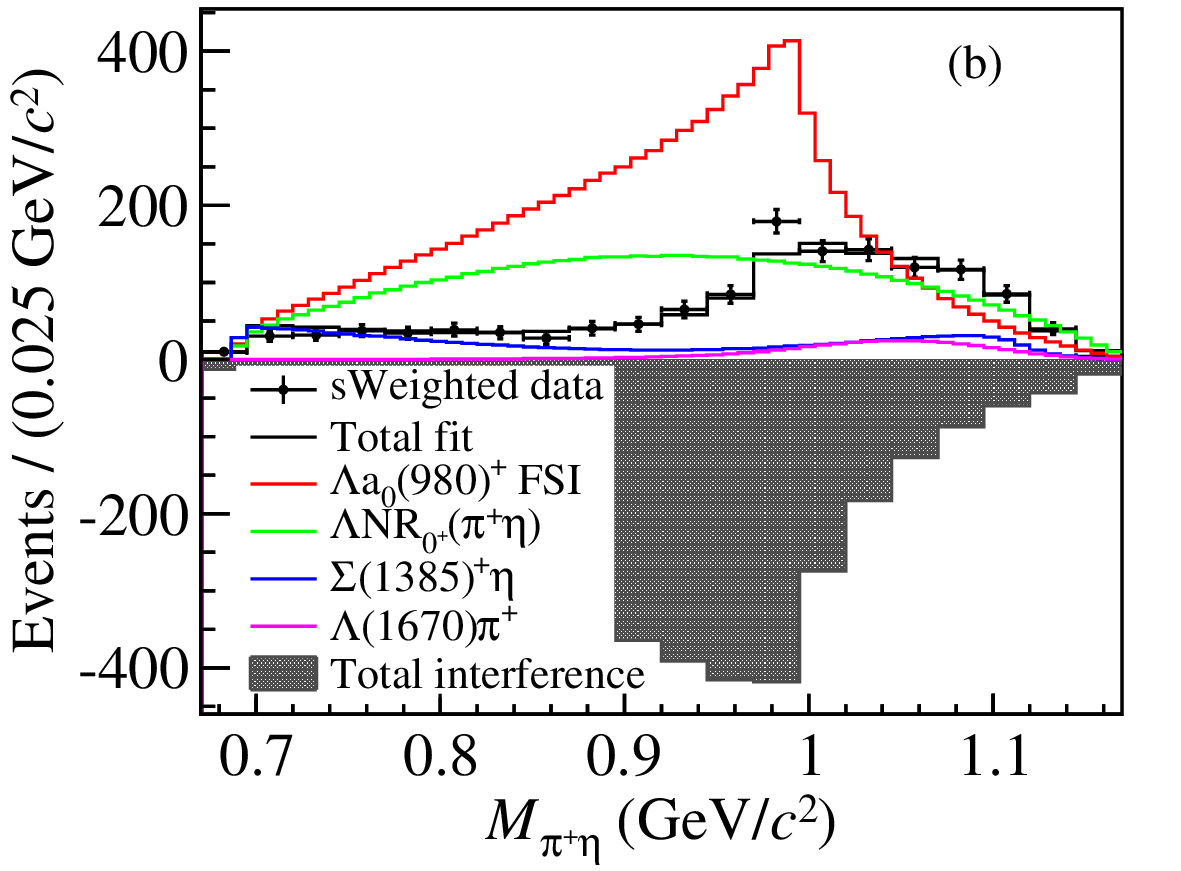}
  	\includegraphics[width=0.32\textwidth]{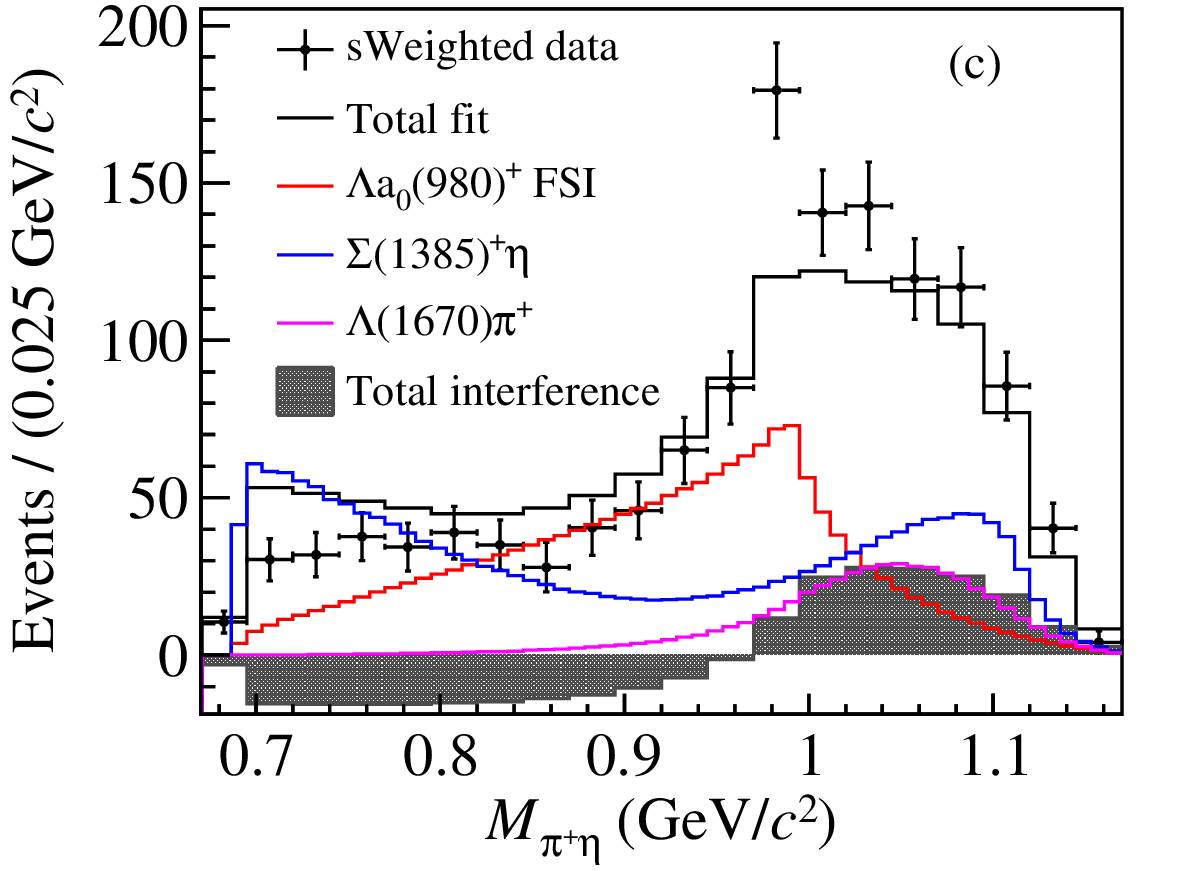}
	\caption{Projections of the fit results using FSI model in the $M_{\Lambda\eta}$ and $M_{\pip\eta}$ spectra. Points with error bars are sWeighted data at all energy points. The curves in different colors are different components.\label{fig:fit_mass_FSI}}
\end{figure}

Alternative PWA fits are performed by replacing the RBW and Flatt\'{e} model with the FSI model~\cite{suppleWang:2022nac}. Figure~\ref{fig:fit_mass_FSI} (a) shows the projection in the $M_{\Lambda\eta}$ spectrum utilizing a $\Lambda(1670)$ FSI model. Figure~\ref{fig:fit_mass_FSI} (b) and (c) show the projections in the $M_{\pip\eta}$ spectrum exploiting the $\a980$ FSI model with and without including a NR component. It is found that if NR and $\a980$ FSI are both included, the interference is very large. However, an $\a980$ FSI without NR cannot describe the $M_{\pip\eta}$ spectrum well.

\section{Systematic uncertainties in PWA}

The systematic uncertainties of the observables in the PWA are now described. The following sources are considered: fixed parameters, barrier radius, additional resonances,  $\lcp$ polarization, fit methods, data-MC differences, and background modeling. All the systematic uncertainties are listed in Table~\ref{tab:PWA_sys}.

\begin{table}[htbp]
	\centering
	\caption{Systematic uncertainties (in units of corresponding statistical uncertainties. For $\alpha_{\Lambda\a980}$, the symmetric statistical uncertainty 0.12 is used.) on the FFs of $\a980,\sgm1385$ and $\lmd1670$, and the corresponding decay asymmetry parameters $\alpha_{\Lambda\a980},\alpha_{\sgm1385\eta}$ and $\alpha_{\lmd1670\pip}$. The total systematic uncertainties are obtained by summing up all contributions in quadrature.\label{tab:PWA_sys}}
    \small
	$\begin{array}{ccccccc}
	\hline \hline 
	& \mathrm{FF}_{\a980} & \mathrm{FF}_{\sgm1385} & \mathrm{FF}_{\lmd1670} & \alpha_{\Lambda\a980} & \alpha_{\sgm1385\eta} & \alpha_{\lmd1670\pip} \\
	\hline 
	\text{Fixed parameters} & 0.09 & 0.10 & 0.32 & 0.08 & 0.05 & 0.11\\
	\text{Barrier radius} & 0.16 & 0.08 & 0.07 & 0.11 & 0.09 & 0.03\\
	\text{Additional resonances} & 0.21 & 0.21 & 0.25 & 0.14 & 0.12 & 1.19\\
	\lcp \text{ polarization} & 0.05 & 0.05 & 0.02 & 0.06 & 0.02 & 0.02\\
	\text{Fit methods} & 0.07 & 0.07 & 0.05 & 0.09 & 0.06 & 0.11\\
	\text{Data-MC differences} & 0.09 & 0.01 & 0.15 & 0.04 & 0.08 & 0.09\\
	\text{Background modeling} & 0.06 & 0.06 & 0.02 & 0.61 & 0.19 & 0.30\\
	\hline
	\text{Total} & 0.31 & 0.27 & 0.44 & 0.66 & 0.27 & 1.24\\
	\hline \hline
	\end{array}$
\end{table}

{\it Fixed parameters.} The uncertainties due to fixed resonance parameters and the $\Lambda$ decay asymmetry parameter are estimated by varying the fixed values within $\pm1\sigma$ and re-fitting the data. The quadratic sums of the largest variations from each parameter are assigned as systematic uncertainties.

{\it Barrier radius.} The systematic uncertainties of the barrier radius $d$ is determined by setting $d$ to the alternative values derived from Ref.~\cite{suppleBESIII:2022udq}, $d=0.53~\mathrm{fm}$ and $d=1.16~\mathrm{fm}$, which is a $\pm 1\sigma$ region. The fit procedure is repeated, and the largest variations are assigned as systematic uncertainties.

{\it Additional resonances.} The systematic uncertainties associated with additional resonances are considered by adding an additional resonance, the $\Sigma(1660)^+$, which, at $4.3\sigma$, is the most significant among the unused resonances. The fit procedure is repeated, and the changes on the fit results are assigned as the systematic uncertainties.

{\it $\lcp$ Polarization.} To estimate the systematic uncertainties due to the $\lcp$ polarization, the fixed parameters, $i.e.$, $\alpha_0$ and $\Delta_0$, are varied by $\pm 1\sigma$, and the fit procedure is repeated. The largest variations are added in quadrature and assigned as systematic uncertainties.

{\it Fit method.} To estimate the potential bias effect due to fit method, we use the results of pull distribution check. For those parameters which are corrected, the uncertainties on the corrections are propagated to obtain systematic uncertainties, and for those parameters which are not corrected, the deviations from 0 (1) in mean (width) are propagated. The total systematic uncertainties are the quadrature sum of mean and width effects.  

{\it Data-MC differences.} To estimate the systematic uncertainties due to the difference between the MC-determined efficiency and the true one, the effects from tracking and PID of the $\pi^\pm$ candidates, and the reconstruction of the $\Lambda$ and $\eta, \piz$ candidates are considered. The reconstruction efficiency differences between data and MC simulations have been studied, for example by using $J/\psi\to p\bar{p}\pip\pim$ control samples for $\pi^\pm$ tracking and PID, $\psi(3686)\to J/\psi\piz\piz$ and $e^{+}e^{-}\to\omega\piz$ control samples for $\eta, \piz$ reconstruction, and $J/\psi\to\bar{p}K^+\Lambda$ for $\Lambda$ reconstruction. The correction factors $w=\varepsilon_{\mathrm{Data}} / \varepsilon_{\mathrm{MC}}$ are assigned as the weighting factors for the PHSP signal MC sample, and the fit procedure is repeated. The resulting variations on the final results are considered as systematic uncertainties.

{\it Background modeling.} To estimate systematic uncertainties due to background description, we change from the sWeight method to a sideband method, in which the events in the $M_{\mathrm{BC}}$ sideband region, $[2.20,2.27]\,\gevcc$, are used to model the background shape in the $M_{\mathrm{BC}}$ signal regions shown in Table~\ref{tab:BDTG_mbcfit}. The fit procedure is repeated, and the changes in the fit results are taken into account as systematic uncertainties.

\section{Fit results for the BF measurement}

\begin{figure}[htbp]
	\centering
    \includegraphics[width=0.24\textwidth]{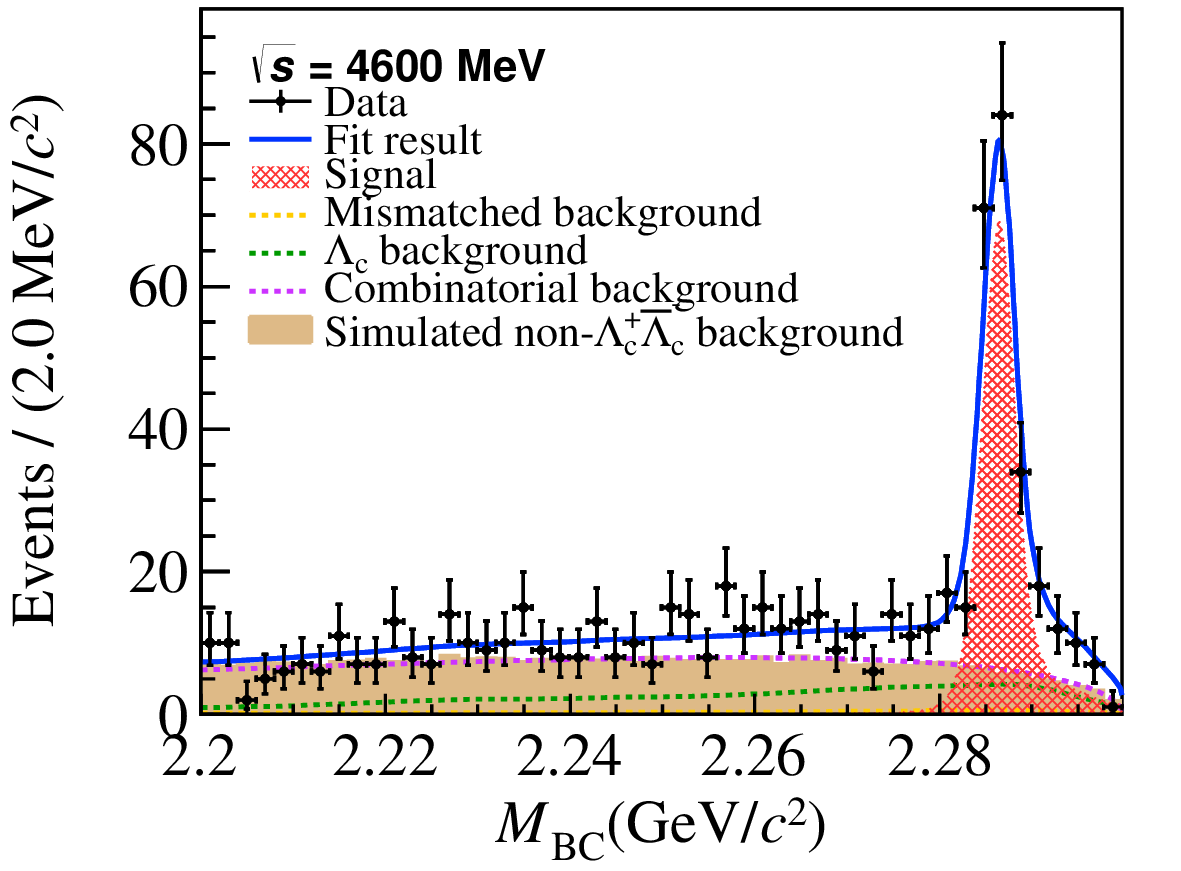}
    \includegraphics[width=0.24\textwidth]{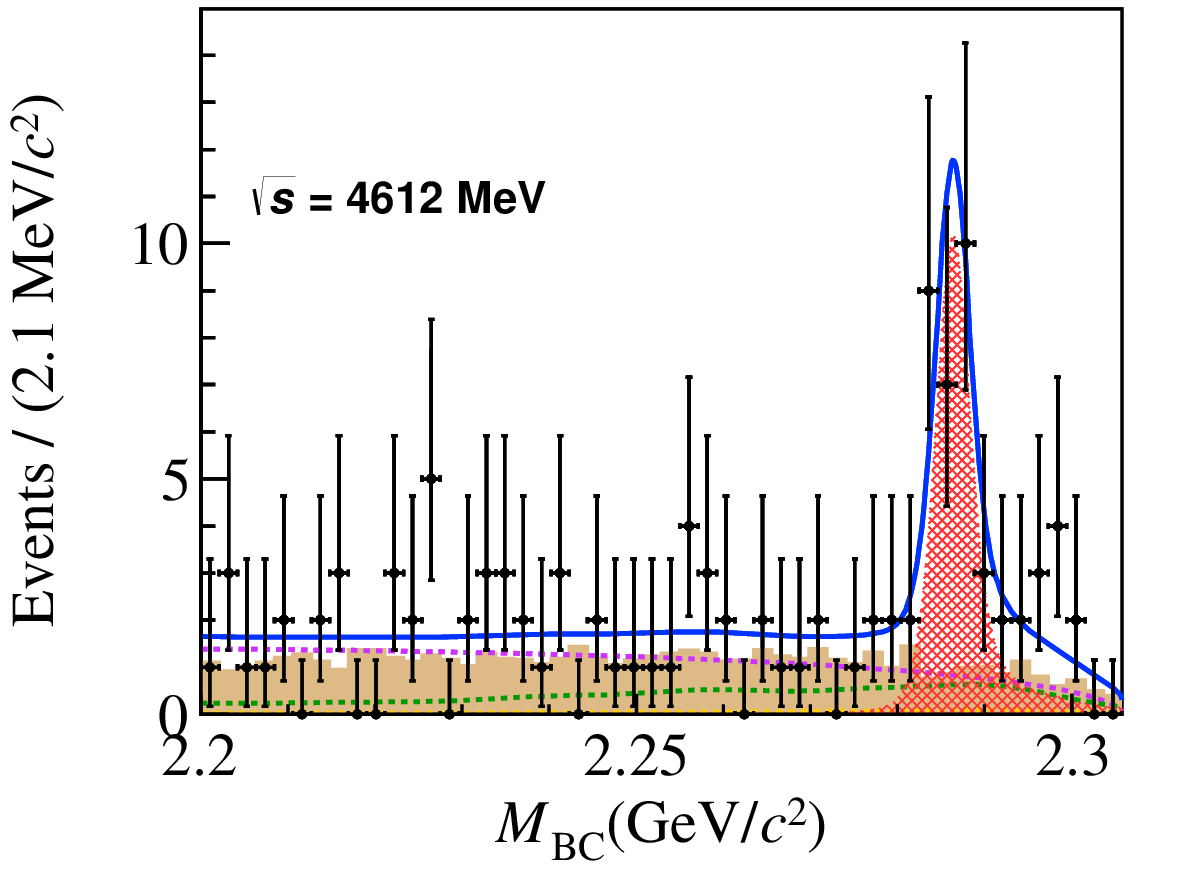}
    \includegraphics[width=0.24\textwidth]{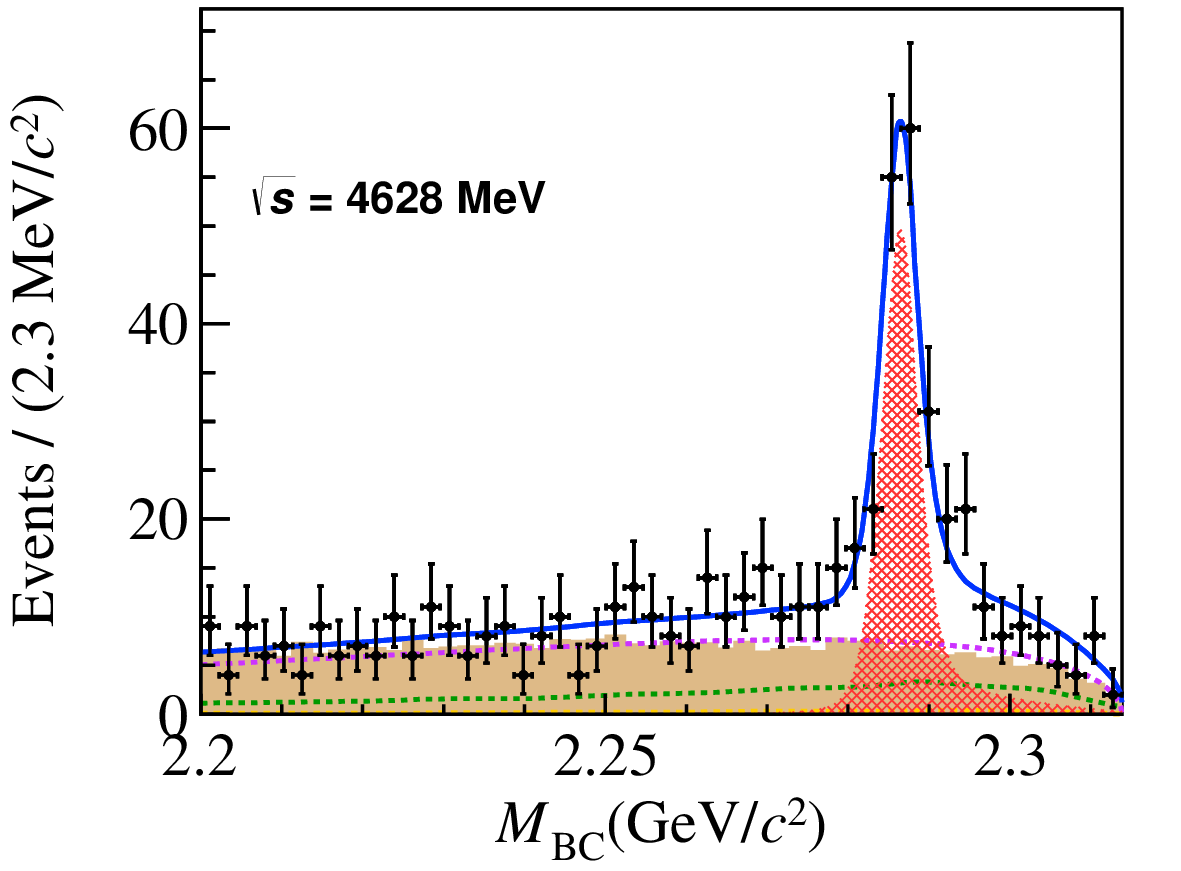}
    \includegraphics[width=0.24\textwidth]{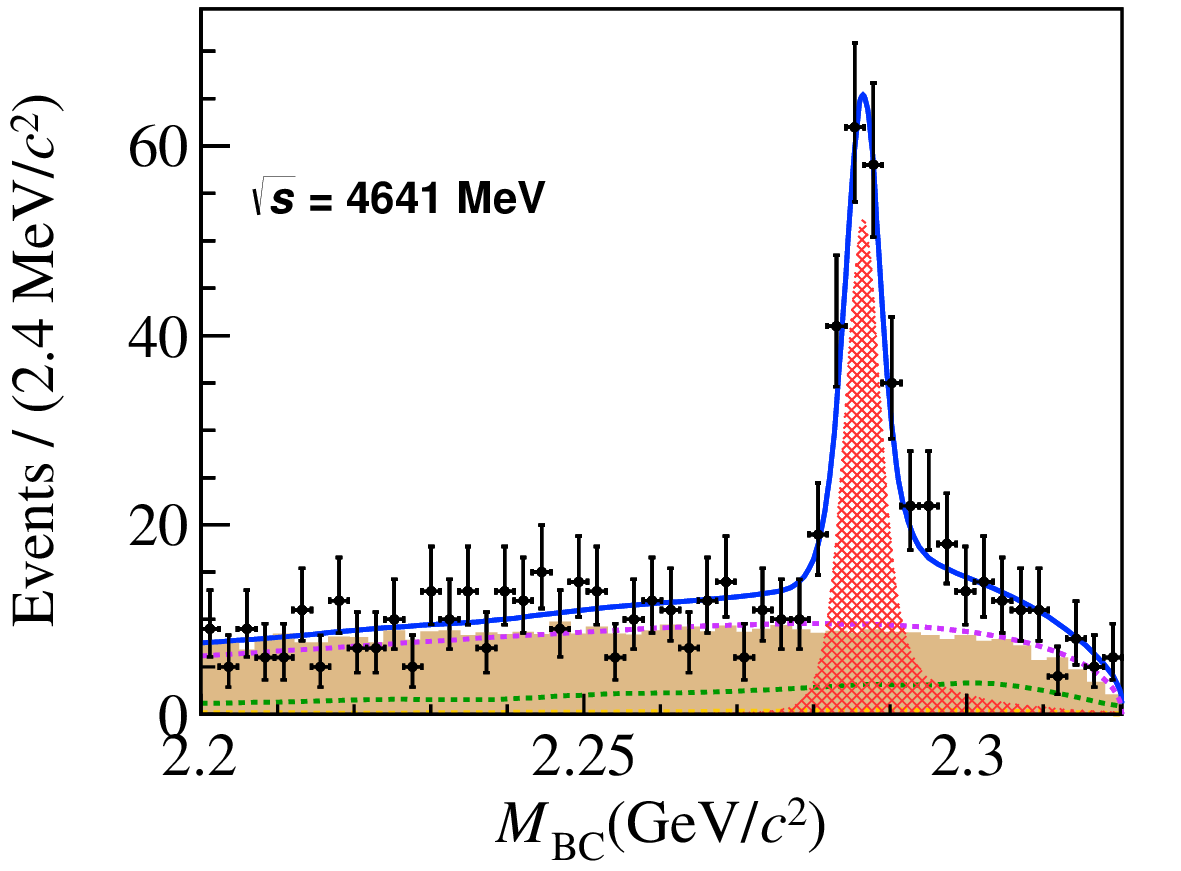}
    \includegraphics[width=0.24\textwidth]{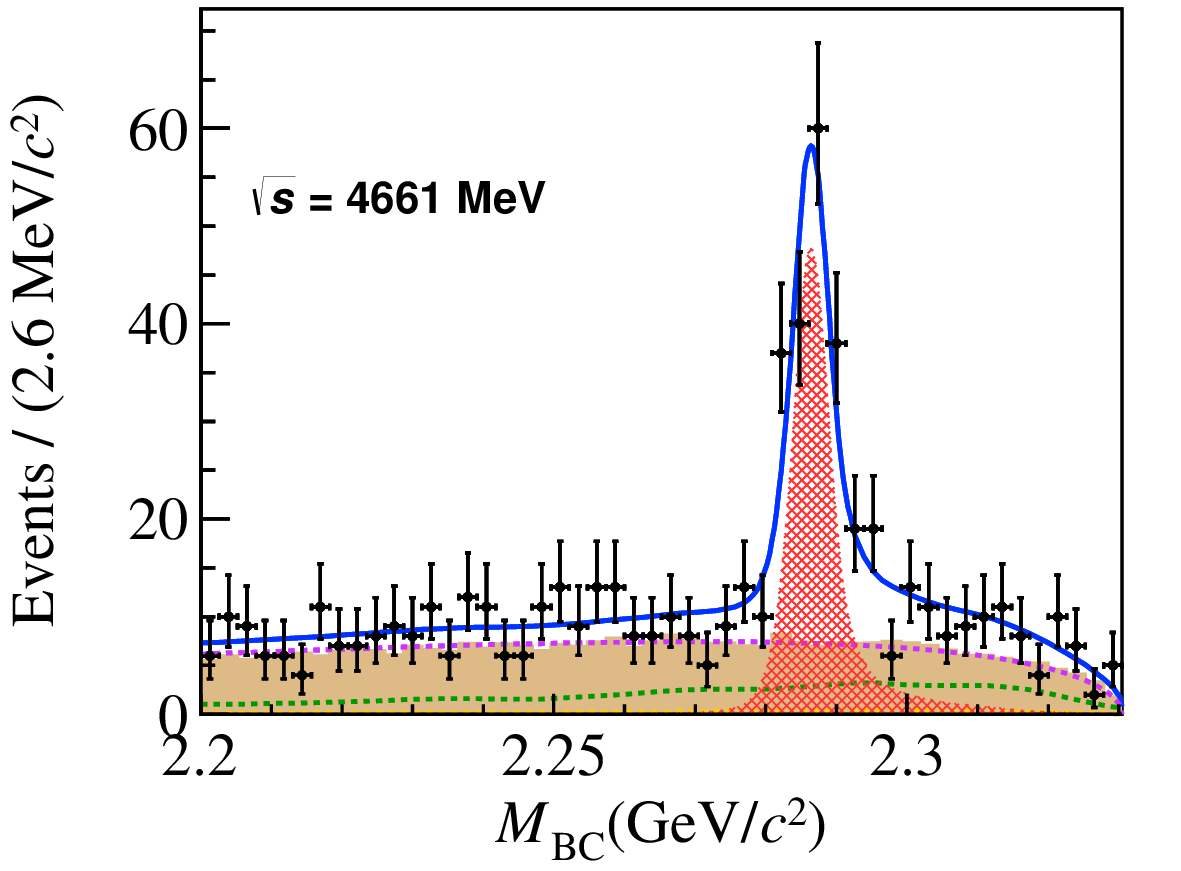}
    \includegraphics[width=0.24\textwidth]{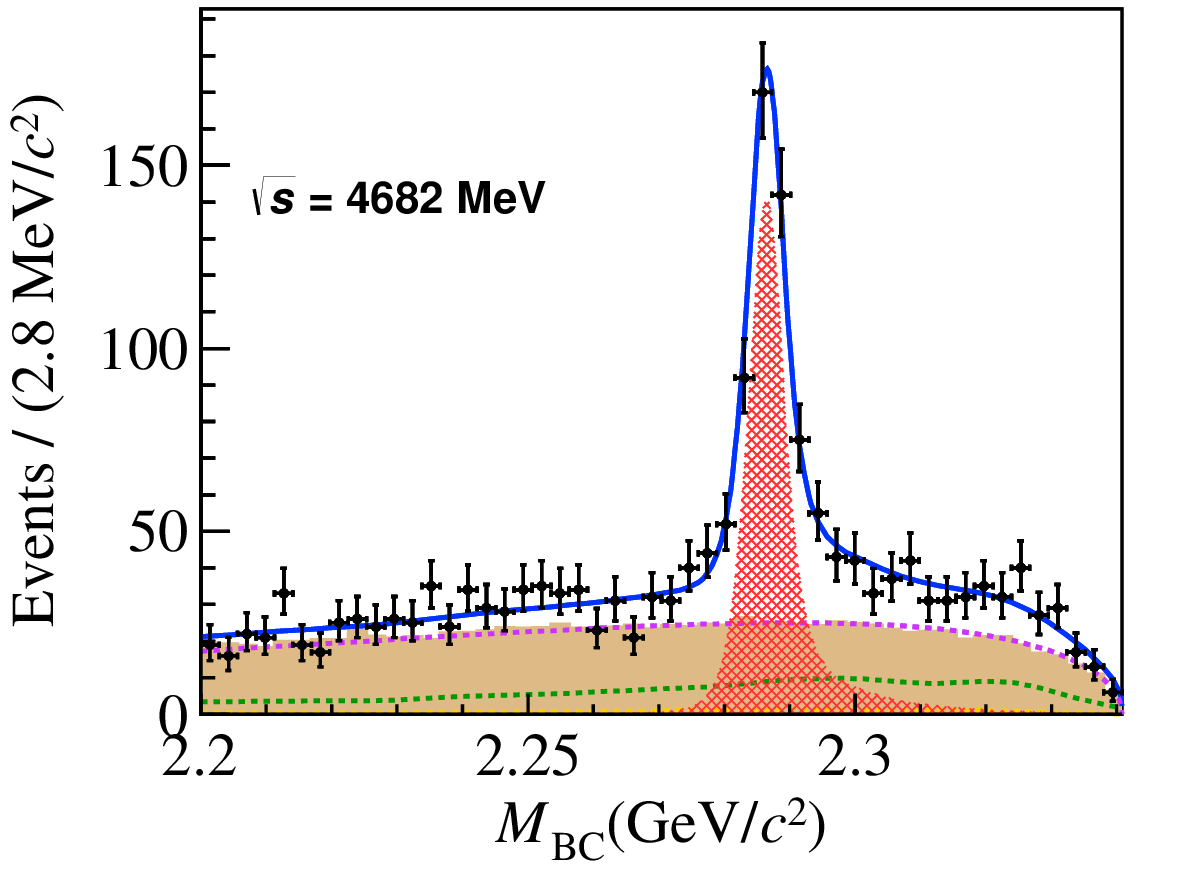}
    \includegraphics[width=0.24\textwidth]{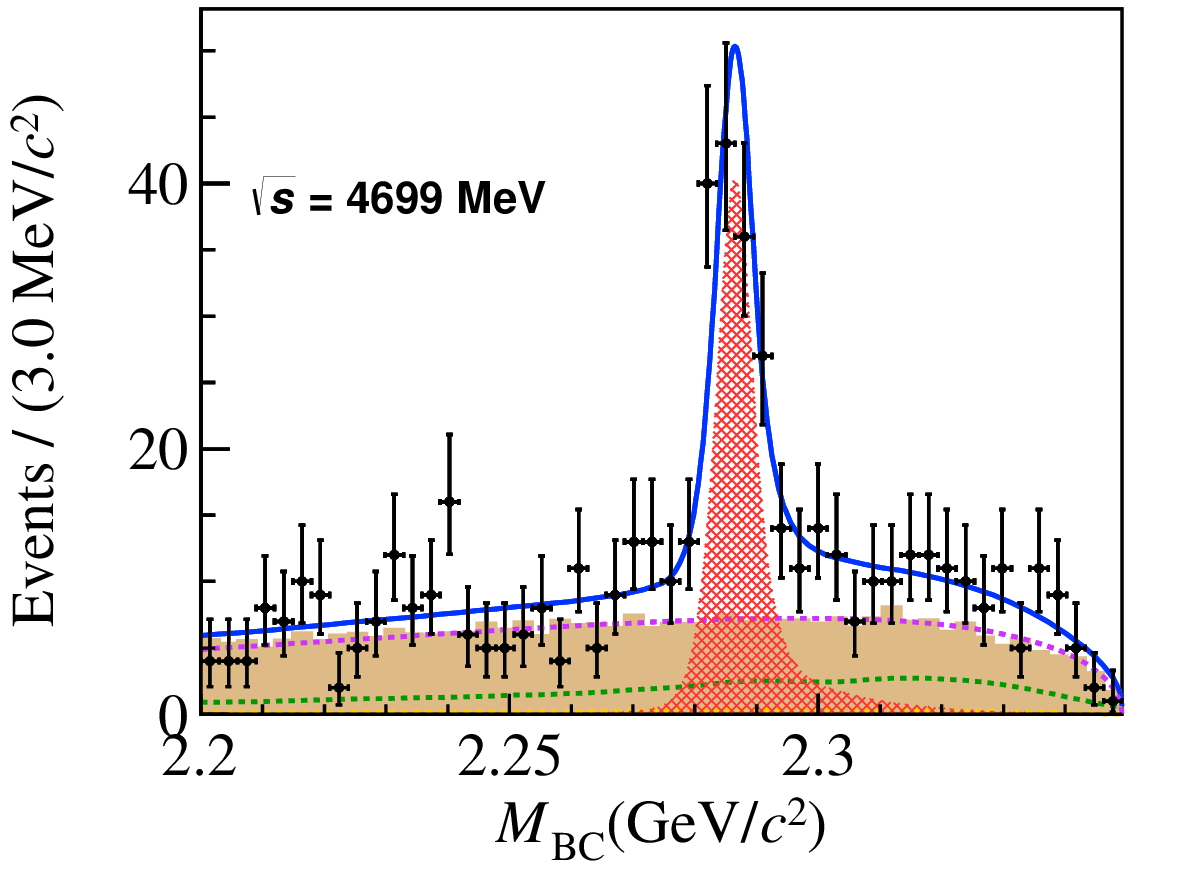}
    \includegraphics[width=0.24\textwidth]{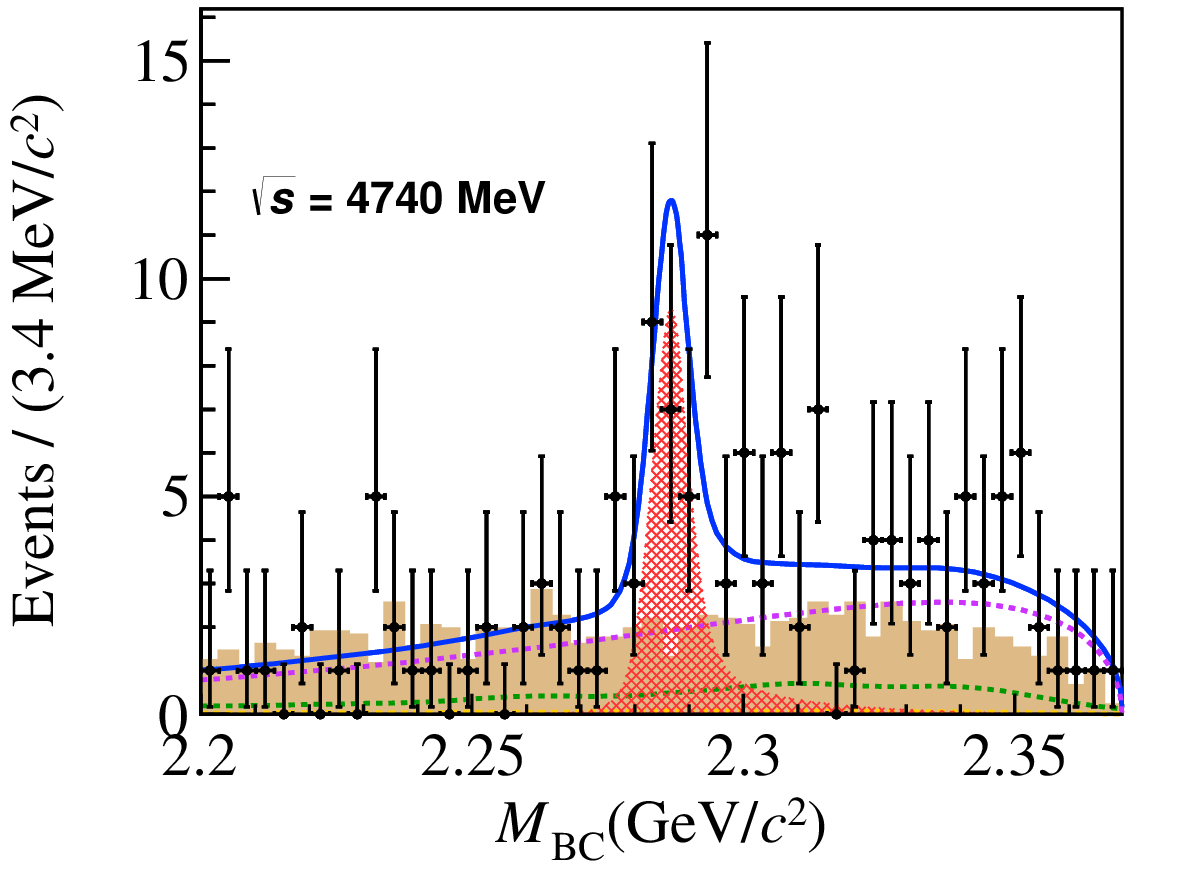}
    \includegraphics[width=0.24\textwidth]{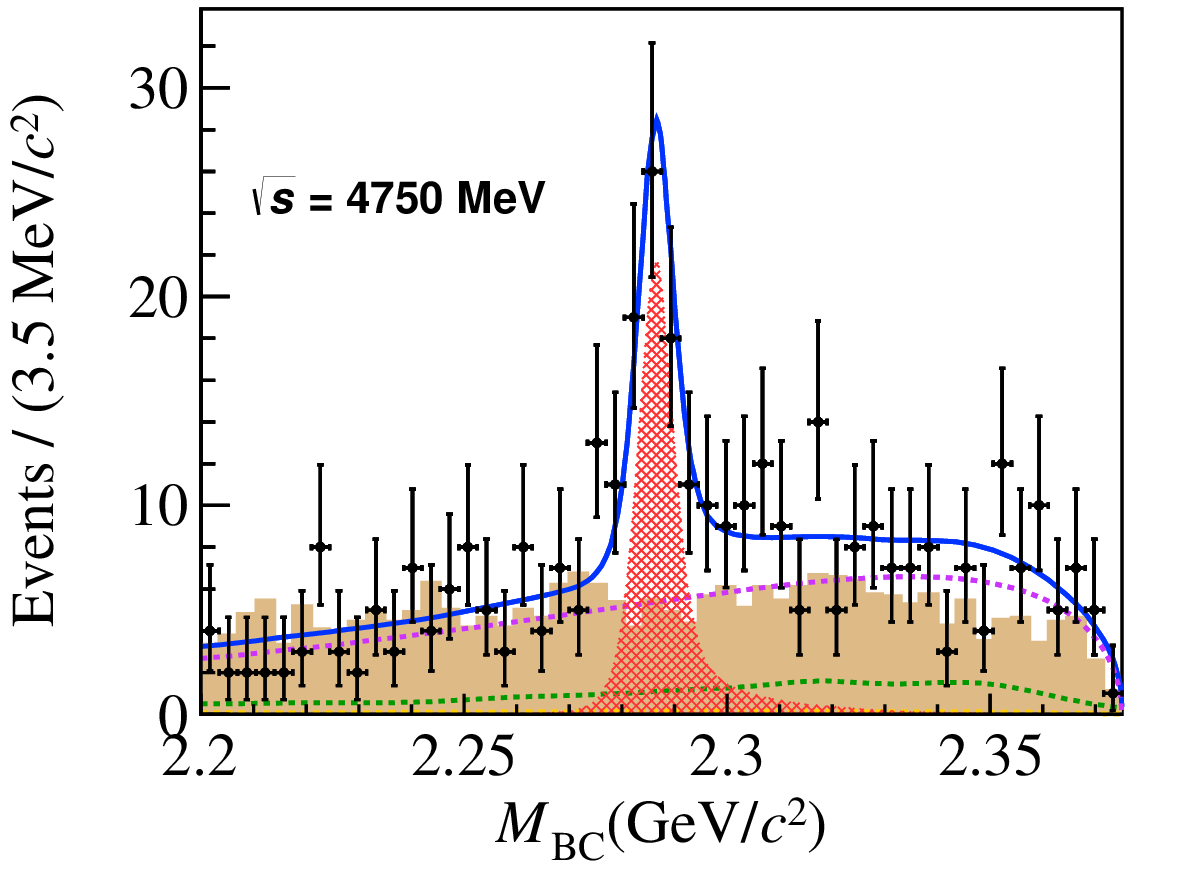}
    \includegraphics[width=0.24\textwidth]{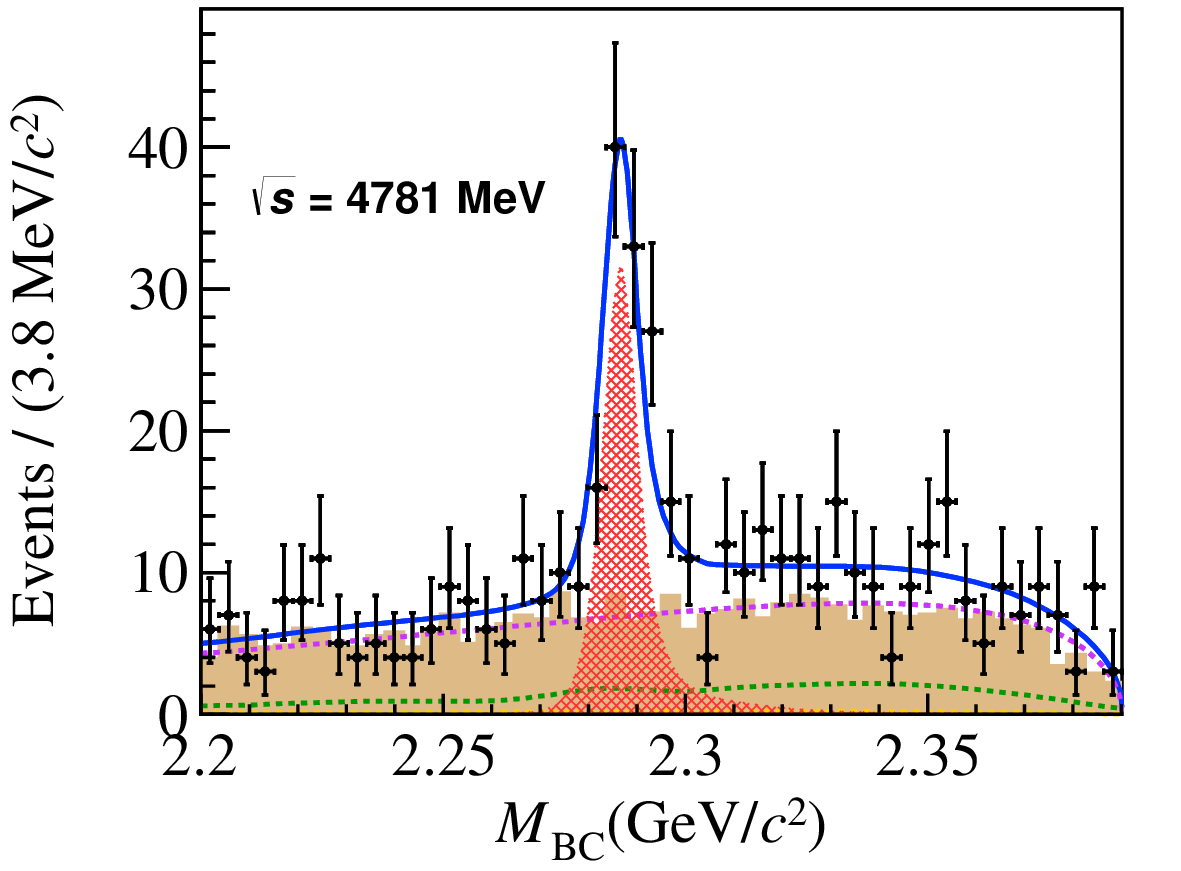}
    \includegraphics[width=0.24\textwidth]{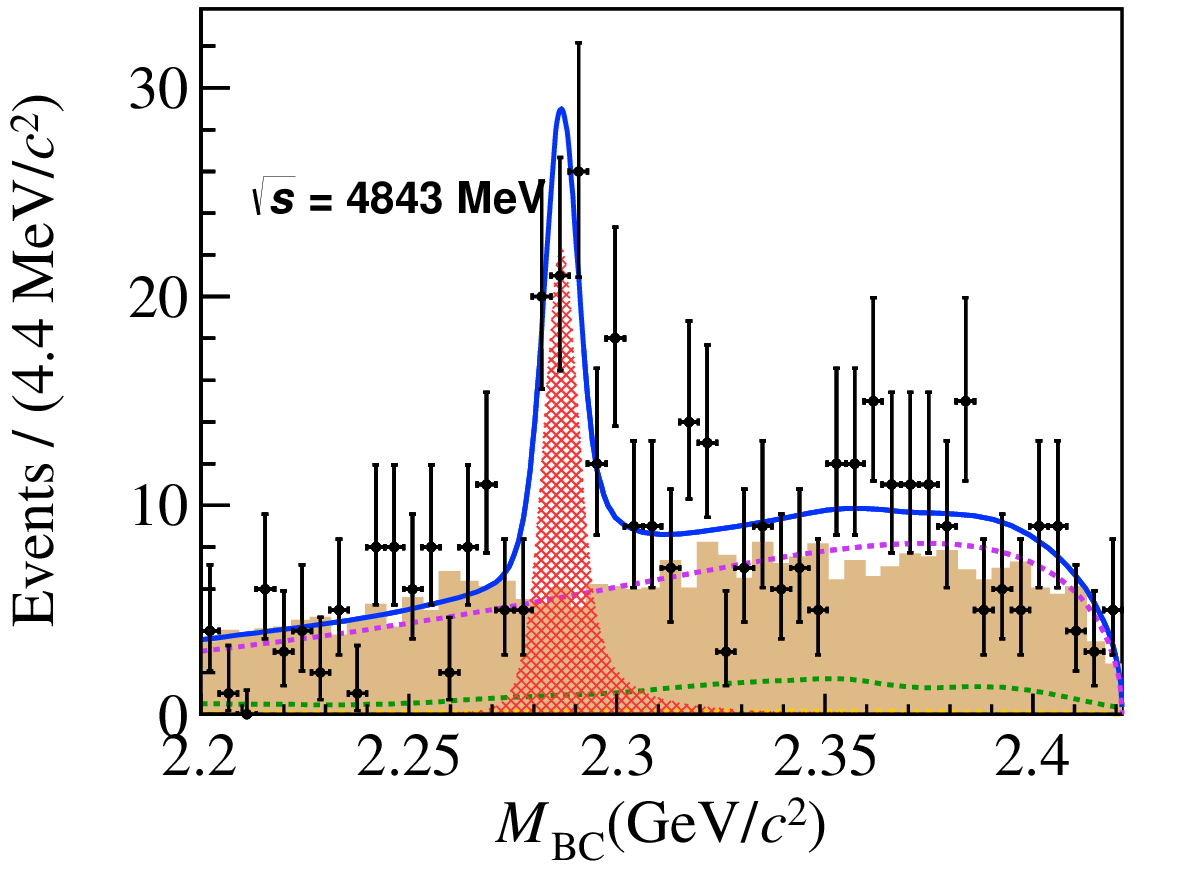}
	\caption{The simultaneous fit to the $M_{\rm BC}$ distributions combined from the $\eta\to\gamma\gamma$ and $\eta\to\pip\pim\piz$ channels at different energy points. The points with error bars are data, the brown solid histograms are MC-simulated background derived from the inclusive MC sample excluding $\lcp\lcm$, the red grid histograms are signal, the orange dashed lines are mismatched background, the green dashed lines are $\Lambda_c$ decay background, the violet dashed lines are combinatorial background shapes, and the blue lines are total fit curves.\label{fig:SimulFit_BF}}
\end{figure}

The average detection efficiency is defined as
\begin{equation}
    \varepsilon = \frac{ \sum_i N_{\lcp\lcm}^i \varepsilon^i }{ \sum_i N_{\lcp\lcm}^i },
\end{equation}
where $N_{\lcp\lcm}^i$ and $\varepsilon^i$ are shown in Table~\ref{tab:BF_Fit_result}. The average detection efficiencies are $(13.73\pm0.02)\%$ and $(4.83\pm0.01)\%$ for the $\eta\to\gamma\gamma$ and $\eta\to\pip\pim\piz$ channels, respectively. The result of the fit to the $M_{\mathrm{BC}}$ distribution combined from the $\eta\to\gamma\gamma$ and $\eta\to\pip\pim\piz$ channels at every energy point is shown in Fig.~\ref{fig:SimulFit_BF}.

\begin{table*}[htbp]
	\centering
	\caption{
	The results of the simultaneous fit to different energy points, along with the numbers of $\lcp\lcm$ pairs ($N_{\lcp\lcm}$), detection efficiencies ($\varepsilon$), the ratios of matched events to mismatched signal yields ($r$) and the signal yields ($N_{\mathrm{sig}}$) in the $\eta\to\gamma\gamma$ and $\eta\to\pip\pim\piz$ channels. The uncertainties are statistical only.\label{tab:BF_Fit_result}}
    \small
	$\begin{array}{c|cccc|ccc}
	\hline \hline 
    \multirow{2}{*}{\text {Sample}} & \multirow{2}{*}{\textit{$N_{\lcp\lcm}$}} & \multicolumn{3}{c}{\eta\to\gamma\gamma} & \multicolumn{3}{c}{\eta\to\pip\pim\piz} \\
    \cline{3-8}
    & & \varepsilon~(\%) & r~(\%) & N_{\mathrm{sig}} & \varepsilon~(\%) & r~(\%) & N_{\mathrm{sig}} \\
	\hline 
	4600 & 99245 \pm 3667 & 15.69\pm0.05 & 5.41\pm0.08 & 152.1\pm5.2 & 5.76\pm0.03 & 20.50\pm0.29 & 32.3\pm1.1\\
	4612 & 17434 \pm 833 & 14.65\pm0.08 &  5.69\pm0.14 & 24.9\pm0.9 & 5.02\pm0.05 &  17.87\pm0.45 & 4.9\pm0.2\\
	4628 & 89286 \pm 3348 & 14.44\pm0.05 &  5.70\pm0.09 & 125.9\pm4.3 & 4.71\pm0.03 &  17.48\pm0.30 & 23.8\pm0.8\\
	4641 & 95435 \pm 3521 & 14.11\pm0.05 &  5.96\pm0.09 & 131.6\pm4.5 & 4.79\pm0.03 &  18.15\pm0.30 & 25.8\pm0.9\\
	4661 & 91644 \pm 3379 & 13.83\pm0.05 &  6.09\pm0.10 & 123.8\pm4.2 & 4.59\pm0.03 &  16.36\pm0.29 & 23.7\pm0.8\\
	4682 & 278621 \pm 9738 & 13.43\pm0.04 &  6.06\pm0.07 & 365.4\pm12.5 & 4.86\pm0.02 &  20.28\pm0.22 & 76.5\pm2.6\\
	4700 & 84348 \pm 3266 & 13.21\pm0.05 &  6.27\pm0.10 & 108.8\pm3.7 & 4.39\pm0.03 &  16.30\pm0.29 & 20.9\pm0.7\\
	4740 & 19845 \pm 1052 & 13.20\pm0.08 &  6.98\pm0.17 & 25.6\pm0.9 & 4.36\pm0.05 &  15.66\pm0.46 & 4.9\pm0.2\\
	4750 & 45086 \pm 1846 & 12.58\pm0.08 &  6.59\pm0.17 & 55.4\pm1.9 & 4.62\pm0.05 &  16.80\pm0.46 & 11.8\pm0.4\\
	4781 & 61428 \pm 2426 & 12.90\pm0.05 &  6.37\pm0.10 & 77.4\pm2.6 & 4.76\pm0.03 &  18.19\pm0.30 & 16.5\pm0.6\\
	4843 & 45426 \pm 1866 & 12.12\pm0.05 &  6.82\pm0.11 & 53.8\pm1.8 & 4.56\pm0.03 &  16.92\pm0.29 & 11.7\pm0.4\\
	\hline
	\hline
	\end{array}$
\end{table*}

\section{Systematic uncertainties in the BF measurement}

The sources of systematic uncertainties are summarized in Table~\ref{tab:BF_sys}; their sum in quadrature is taken as the total systematic uncertainty.

\begin{table}[htbp]
	\centering
	\caption{Systematic uncertainties in the BF measurement.\label{tab:BF_sys}}
	$\begin{array}{cc}
	\hline \hline 
	\text{Sources} & \mathcal{B}(\lcp\to\lmdetapi)(\%) \\
	\hline 
	\text{Tracking} & 0.9\\
	\text{PID} & 0.3\\
	\Lambda \text{ reconstruction} & 2.6\\
	\eta \text{ reconstruction} & 1.0\\
	\text{BDTG cut} & 1.1\\
	\text{Signal model} & 2.7\\
	\text{Fit model} & 0.9\\
	\mathcal{B}_{\mathrm{inter}} & 0.9\\
	N_{\lcp\lcm} & 3.9\\
	\text{MC statistics} & 0.4\\
	\hline
	\text{Total} & 5.7\\
	\hline \hline
	\end{array}$
\end{table}

\begin{itemize}
	\item 	\emph{Tracking, PID, $\Lambda$ and $\eta$ reconstruction}. We use $J/\psi \to p\bar{p}\pip\pim$ control samples to determine the pion tracking efficiency, $J/\psi\to \bar{p} K^{+} \Lambda, p K^{-} \bar{\Lambda}$ to determine the $\Lambda$ reconstruction efficiency, and $\psi(3686)\to J/\psi \, \piz\piz$ and $e^{+}e^{-}\to\omega\piz$ control samples to determine the $\eta$ reconstruction efficiency. The detection efficiency is re-weighted, and an alternative simultaneous fit is performed. The difference between the nominal and alternative fit results is taken as the uncertainty.
	
	\item	\emph{BDTG cut}. The systematic uncertainty due to the BDTG cut is studied with a $\lcp \to \Lambda \pip \piz$ control sample. The difference between the BF before applying the BDTG cut and that after the cut, 1.1\%, is assigned as the systematic uncertainty.
	
	\item	\emph{Signal model}. The systematic uncertainty due to signal model is assigned to be 2.7\%, by varying the amplitude analysis model parameters according to the error matrix and re-weighting the detection efficiency.
	
	\item	\emph{Fit model}. To estimate the systematic uncertainty due to the fit model, we re-weight the signal shape according to the error matrix and also vary the fixed background numbers by $\pm1\sigma$. The quadrature sum of the difference between nominal and alternative fit results, 0.9\%, is taken as the uncertainty.
	
	\item	\emph{$\mathcal{B}_{\mathrm{inter}}$}. The systematic uncertainty due to the $\mathcal{B}_{\mathrm{inter}}$ factor is estimated by varying it by $\pm 1\sigma$~\cite{supplepdg2023}. The deviation from the nominal fit result, 0.9\%, is assigned as the systematic uncertainty.
	
	\item	\emph{$N_{\lcp\lcm}$}. The systematic uncertainty due to $N_{\lcp\lcm}$ is estimated by varying this yield by $\pm1\sigma$. The change of the fit result, 3.9\%, is taken as the systematic uncertainty.
	
	\item	\emph{MC statistics}. The uncertainty related to MC statistics is estimated by varying the detection efficiencies by $\pm1\sigma$. The change of the nominal fit result, 0.4\%, is taken as the systematic uncertainty.
	
\end{itemize}

\end{appendices}

\end{document}